\documentclass[aps,prb,reprint,twocolumn,superscriptaddress,preprintnumbers,nofootinbib]{revtex4}

\usepackage[english]{babel}
\usepackage{amsthm}
\usepackage{amsmath}
\usepackage{graphicx}
\usepackage{slashed}
\usepackage{amssymb}
\usepackage{float}
\usepackage[utf8]{inputenc}
\usepackage[T1]{fontenc}
\usepackage[colorlinks=True, citecolor=blue, urlcolor=blue, linkcolor=blue]{hyperref} % Required for inserting images
\usepackage{color}

\usepackage[normalem]{ulem}
\usepackage{textgreek}

\definecolor{pink}{rgb}{1,0.5,0.5}
\definecolor{violet}{rgb}{1,0,1} % color values Red, Green, Blue
\definecolor{red}{rgb}{1,0,0}
\definecolor{yellow}{rgb}{0.7,1,0}
\definecolor{orange}{rgb}{1,0.5,0}
\definecolor{white}{rgb}{1,1,1}
\definecolor{blue}{rgb}{0,0,1}
\definecolor{cyan}{rgb}{0,1,1}

\newcommand{\om}{\omega}

\newcommand{\beq}{\begin{equation}}
\newcommand{\eeq}{\end{equation}}
\newcommand{\bea}{\begin{eqnarray}}
\newcommand{\eea}{\end{eqnarray}}

\allowdisplaybreaks

\begin{document}

\title{Quantum theory of elastic strings and the thermal conductivity of glasses}

\author{Fernando Lund}
\email{flund@dfi.uchile.cl}
\affiliation{Departamento de F\'{i}sica, Facultad de Ciencias F\'{i}sicas y Matem\'{a}ticas, Universidad de Chile, Santiago, Chile}

\author{Bruno Scheihing-Hitschfeld}
\email{bscheihi@kitp.ucsb.edu}
\affiliation{Kavli Institute for Theoretical Physics, University of California, Santa Barbara, California 93106, USA}

\date{\today}

\begin{abstract}
    We study the thermal conductivity of amorphous solids by constructing a continuum model whose degrees of freedom are propagating vibrational modes (phonons) and extended Volterra dislocation line defects with their own vibrational degrees of freedom which do not propagate in space. Our working assumption is that these additional degrees of freedom account for the ``boson peak'' that is observed in glassy materials. This identification allows us to obtain the length distribution of dislocations from experimental data of the boson peak for each material, which we use as input to calculate the phonon self-energy in a quantum field theory framework using that the phonon-dislocation interaction is given by the Peach-Koehler force. The tail of the distribution for long dislocations is consistent with an $L^{-5}$ power law. Our results show that this power law yields a linear rise in the thermal conductivity, as observed in glasses at low temperatures. We then consider two approaches to describe thermal conductivity data quantitatively. In the simplest approach we only keep the low-frequency behavior of the phonon self-energy with one free parameter, plus an adjustable UV cutoff. In the more realistic approach we keep the full frequency dependence of the phonon self-energy as dictated by the phonon-dislocation interaction plus an additional contribution due to scattering with point defects, with a cutoff set by the typical interatomic spacing of the material. We obtain a satisfactory description of thermal conductivity data with both approaches. We conclude by discussing prospects to test the predictive power of this model. 
\end{abstract}

\maketitle

\section{Introduction}
Amorphous materials, as opposed to crystals,  do not have long range translational order at the atomic scale. Consequently there is no Bloch theorem to understand the electronic states, and no Brillouin zone to classify the normal modes of vibration near equilibrium atomic positions. The glass transition itself (i.e., the dramatic increase of the viscosity of a supercooled liquid as it is cooled) is a phenomenon shrouded in mystery: in spite of an enormous wealth of experimental data obtained through several decades, centuries even, of research, there is no quantitative understanding comparable to, say, the understanding of equilibrium phase transitions and critical phenomena \cite{Berthier2011,Berthier2023}.  In addition, although from a microscopic point of view there exist a great variety of glasses (oxides, metallic, organic, calchogenides, for example), there are a number of properties that are shared by many of them independently of microstructure, such as the presence of localized normal mode vibrations, the dispersion relation of acoustic waves in the THz regime, and the temperature dependence of the thermal conductivity.  To provide a tool to quantitatively advance in the understanding of glass properties, including their commonalities, is then a major, and long-standing, scientific challenge \cite{Varshneya2019}.

In addition to the extant scientific challenge, it is difficult to exaggerate the importance of advancing our understanding of amorphous materials, given their applications in all walks of life, including the pharmaceutical industry~\cite{Luo2024}, energy management and thermoelectrics~\cite{Baskaran2024}, quantum sensing, communication and computation technologies~\cite{Suleiman2020}, biomaterials for tissue regeneration~\cite{Zhu2024}, data-storage~\cite{Anderson2023}, biomineralization in the life and the geosciences~\cite{Gilbert2022}, and improved optical coatings for gravitational-wave research~\cite{Prasai2023}.

\subsection{Thermal conductivity of glasses}
Of the many properties of amorphous materials that have been extensively studied over decades, the thermal conductivity $\kappa$ of glasses stands out~\cite{Tanguy2023}. Broadly speaking, there is a characteristic behavior over three distinct temperature ranges. Below $T\sim 1$ K, $\kappa \sim T^2$, it can be rationalized in terms of the tunneling two-level system (TLS) model~\cite{Anderson1972,Phillips1972,Carruzzo2020,Carruzzo2021}. This is a simple and general model that allows for quantitative computation, but that has defied to this day a proper link to the atomic structure of glasses. Around $T\sim 10$ K, $\kappa$ is constant (the plateau), a behavior that is qualitatively blamed on an increased scattering of energy carriers but with even less quantitative relationship to the glass microstructure. And for $T$ above 20 K, $\kappa$ increases monotonically, in sharp contrast to the decrease observed in that temperature range for the corresponding crystalline solids. It is this behavior that we address in the present article.

In the absence of a satisfactory physics-mathematical tool, the preferred way to study thermal transport in amorphous materials, particularly above the plateau,  has been through numerical simulation, of which the classic paper by Allen and Feldman~\cite{AllenFeldman1993}  has provided much inspiration, including a conceptual framework in use to this day. In spite of much insight gained from the developments thereof, the need to integrate anharmonic and quantum effects, together with the computational cost, have proved insurmountable obstacles until now to generate numbers that can be compared with experimental results. Recently, Simoncelli et al. have developed a unified theory of thermal transport in crystals and glasses based on a Wigner, rather than a Peierls-Boltzmann, transport equation~\cite{Simoncelli2019,Simoncelli2022} and used it to compute thermal properties of amorphous silica~\cite{Simoncelli2023}, alumina~\cite{Harper2023} and hafnia~\cite{Zeng2025}. In the case of silica~\cite{Simoncelli2023} there is good quantitative agreement between theory and experiment for the thermal conductivity above the plateau from about 80 K up to about 200 K with theory following, but steadily diverging from below, with the qualitative trend up to about 500 K with experimental data for a thin (190 nm) film~\cite{Lee1997} but consistently underestimating, while following the qualitative trend, the bulk experimental results~\cite{Cahill1990}. A related approach has been developed by Baroni et al.~\cite{Isaeva2019,Fiorentino2023a,Fiorentino2023b}. 

The papers mentioned in the previous paragraph address the root of the problem: what do the atoms do in a glass that is responsible for the thermal transport? In the present paper we take an approach from the opposite end of the length scale, the macroscopic scale: We describe a glass as a continuum solid endowed with a population of elastic strings of finite length with ends pinned~\cite{Lund2015,Bianchi2020}. The normal modes of the strings are responsible for the excess of vibrational states observed in glasses in the THz range, so that given a glass and its density of states (measured for example with neutron or X-ray scattering) it is possible to {\it deduce} its thermal conductivity. The continuum solid has phonons labelled by a continuous wave vector, and they interact, through a Peach-Koehler force, with the elastic strings. In addition, in the case of silica, we allow the glass to have point defects whose interaction with the phonons is modeled as that of a simple harmonic oscillator. This basic physics is sufficient to compute the thermal conductivity of {\it any} glass. In the following we carry out this program for amorphous SiO$_2$, As$_2$O$_3$, Ca-K nitrate, and PMMA in the temperature range 30-300 K~\cite{Cahill1987},  glycerol in the temperature range 50-200 K~\cite{Cahill1987} and SiO$_2$ in the temperature range 30-760 K~\cite{Cahill1990}. There is considerably more data available for glycerol and, especially, for SiO$_2$. In both cases the comparison between theoretical results and experimental data is, accordingly, more elaborate.

\section{The string model}
\subsection{Background}
Over the past couple of decades there has emerged evidence, from several numerical simulations of glasses, of the cooperative motion of chains of atoms, that have been called ``strings''.
Cooperative string-like motion has been found for example in numerical simulations of a soft-sphere glass ~\cite{Schober1993}, of super-cooled liquids~\cite{Donati1998,Zhang2015}, and recently it has been found to be linked to the boson peak \cite{Hu2023}. Indirect experimental evidence in favor of an interpretation involving strings of atoms include the relaxation behavior of metallic glasses \cite{Yu2013}, an interpretation for which favorable arguments have been advanced from numerical simulations \cite{Yu2017}, and the anisotropy induced by compression, also of a metallic glass \cite{Concustell2011}. On the theory side, within the random first order transition theory approach to the glass transition, the activated motion between local minima in the energy landscape involves the cooperative motion of string-like clusters of particles \cite{Lubchenko2007}. In this paper we shall look at this string idea from a continuum mechanics point of view.

\subsubsection{The vibrational density of states of glasses and its modeling in terms of elastic strings coupled to phonons through the Peach-Koehler force. Consequences for the behavior of acoustic wave dispersion \cite{Lund2015}.} 

In continuum mechanics, the normal modes of an elastic solid are counted using the classical theory of elasticity. However, an artificial short distance, or  high frequency, cut-off must be introduced, the Debye frequency $\om_D$, to remedy the fact that there is no intrinsic length scale, but the total number of normal modes of any given material is finite. At wavelengths long compared to interatomic spacing, this approach provides a firm foundation to explain all properties of solids that depend on the counting on such modes. If the solid is crystalline, the system is invariant under discrete translations, and a similar counting can also be performed. This counting reduces to the Debye model at long wavelengths, and also provides a firm foundation for properties at shorter wavelengths, down to the size of the unit cell \cite{kittel}

The situation for amorphous solids, without a discrete translation invariance, has long been unsatisfactory. At long wavelengths the situation is well described, as expected, by the Debye model. However, at wavelengths on the order of tens of mean interatomic distances, abundant evidence, from specific heat \cite{Zeller1971}, thermal conductivity \cite{Courtens2001}, Raman scattering \cite{Hehlen2012}, neutron scattering \cite{Bove2005}, and inelastic X-ray scattering measurements \cite{Sette1998}, indicates the existence of normal modes with a frequency distribution that is peaked around 0.1-0.2 $\om_D$. Remarkably, this distribution is qualitatively similar for many such materials, and the details, but not the broad features, depend on external parameters such as temperature, density, pressure, as well as chemical and thermal history \cite{Wischnewski1998,Caponi2007,Inamura2000,Zanatta2010,Monaco2006,Orsingher2010,Niss2007,Hong2008,Caponi2009,Monaco2006}. The existence of this distribution, dubbed the ``boson peak'' (BP), deviates without ambiguity from the distribution for a continuum, at frequencies where the continuum approximation works reasonably well in the case of crystals. Its existence   has generated much research \cite{Parshin1994,Gotze2000,Schirmacher1998,Schirmacher2006,Schirmacher2007,Grigera2003,Ganter2010,Beltukov2013}, but  no satisfactory consensus exists as of yet,  although progress has been achieved through simulations with Lennard-Jones \cite{Monaco2009a}, soft sphere \cite{Maruzzo2013}, power law and Kob-Andersen \cite{Hu2023}  potentials---for a recent review see \cite{Lerner2021}---and through inelastic X-ray scattering experimental studies of SiO$_2$ glasses and crystals\cite{Chumakov2015}. %\fl{\it Additional recent refs: Lerner, Baggioli PRL(2019)} \fl{add Hirota, Lerner Baggioli?}

The behavior of acoustic attenuation as a function of frequency has also been a focal point at THz frequencies.  Acoustic attenuation as a function of frequency $\omega$ behaves like $\om^2-\om^4-\om^2$ as frequency  increases, with the change in slope occuring at boson-peak frequencies \cite{Ruffle2003,Ruffle2006,Monaco2009,Ruta2010,Baldi2010,Baldi2011h,Baldi2011,Ruta2012}: the question naturally comes to mind that boson peak occurence and anomalous sound attenuation should have the same physical origin. In 2015 Lund \cite{Lund2015} provided a possible explanation for this behavior, on the basis of a continuum solid with a random distribution of elastic strings (``Volterra dislocations'') coupled to phonons through the Peach Koehler force. This also explained why, at similar length scales, continuum mechanics works similarly well for crystals and for their amorphous counterparts. This analysis has been elaborated upon by Baggioli and collaborators \cite{Jiang2024a,Jiang2024,Mahajan2025}.

\subsubsection{A success story: Volterra dislocation modeling of glassy Silica and glassy Glycerol; theory, numerics and experiment \cite{Bianchi2020}}

In a further development of the ideas presented in \cite{Lund2015}, Bianchi, Giordano and Lund \cite{Bianchi2020} (a work we will refer to as BGL) computed the density-density correlation function of the continuum-solid-plus-elastic-strings model and the result was successfully compared with experimental data obtained by inelastic x-ray scattering on two prototypical glasses, glycerol and silica. 
{It was also shown that, once the vibrational density of states has been measured, it can be used for {unambiguously} fixing the string length distribution inherent to that glassy system as well as the density-density correlation function. There resulted a fit of impressively good quality for both glycerol and silica in a large wavevector range around the boson peak region. } The model was also able to explain the behavior of the sound velocity as well as the attenuation of the elastic waves, including the $\om^2-\om^4-\om^2$ behavior mentioned in the previous subsection.

An extra bonus of the approach, in which Volterra dislocations couple to phonons through the PK force, is that it was possible to compute the particle motion associated with vibrations of the strings. There resulted a motion with a spatial amplitude similar to the behavior observed for the low energy non-phonic modes in recent numerical simulations \cite{Shimada2018,Lerner2016,Gartner2016,Kapeteijns2018,Lerner2018}. Finally, the string length distribution (which is not a parameter but is determined by the data) is compatible with the reported elastic heterogeneities correlation length for silica \cite{Baldi2013} and dynamical heterogeneities in supercooled liquids \cite{Tracht1998,Berthier2005}. These common properties among significantly different glasses suggest a possible universality, in the sense of their being shared by glasses with very different microstructure.

\subsection{Formalism}

\begin{figure}
    \centering
    \includegraphics[width=0.9\linewidth]{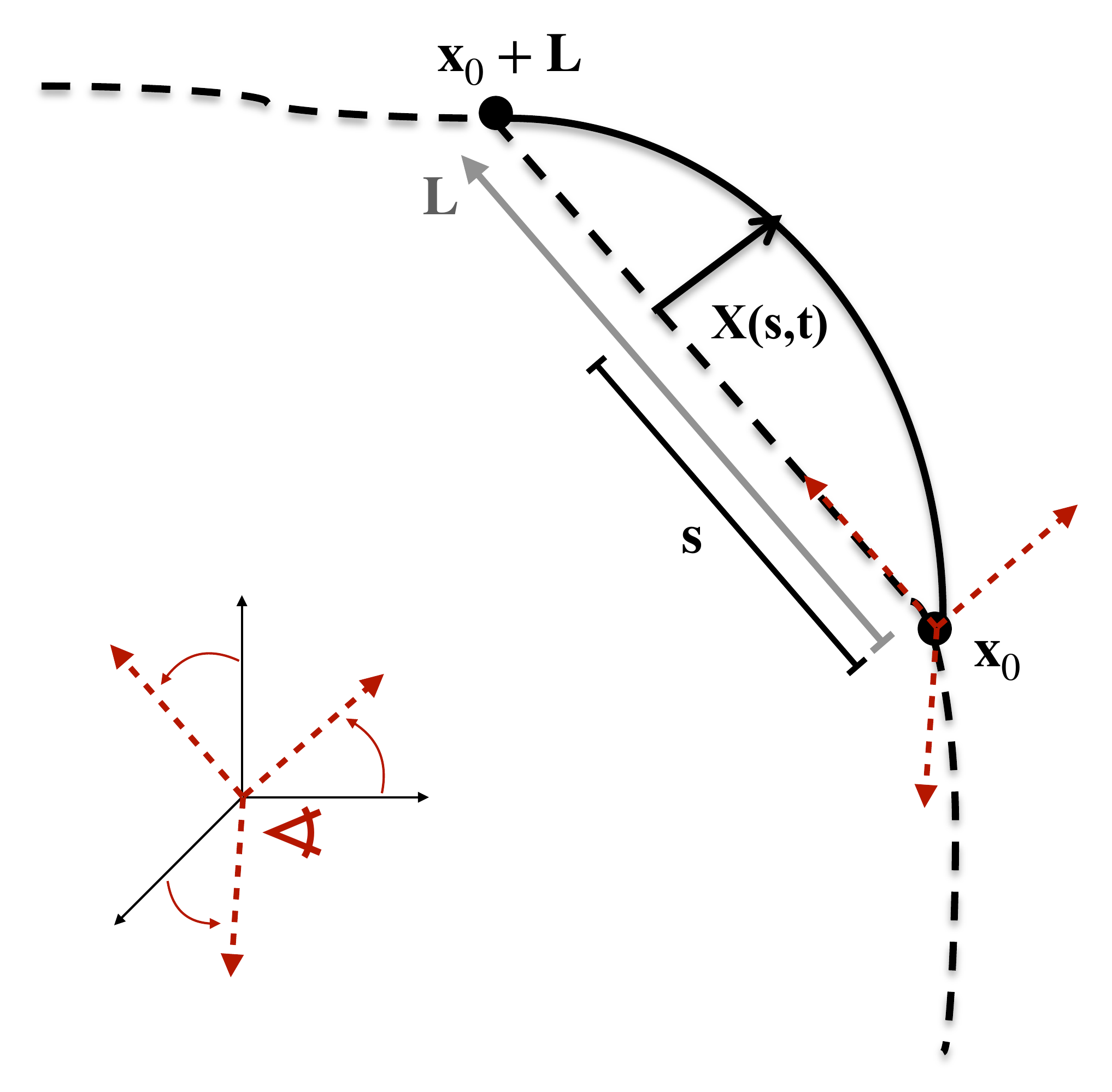}
    \caption{Diagram depicting the various quantities involved in the description of the string that embodies the line defect in an otherwise homogeneous material. The ends of the string are pinned at positions ${\bf x}_0$ and ${\bf x}_0 + {\bf L}$. Its displacement from its equilibrium position is ${\bf X}(s,t)$, and $\sphericalangle$ is the 3D rotation angle that characterizes the orientation of the string relative to a fixed reference frame. All of these quantities (${\bf x}_0, {\bf L}, \sphericalangle$) will be averaged over in the following sections. %\color{red} B: add more labels to the figure!
    }
    \label{fig:dislocation-drawing}
\end{figure}

The model we study in this work is an effective description for the long-wavelength vibrational degrees of freedom of glasses. We consider a homogeneous, isotropic, elastic, continuum of density $\rho$ and elastic constants $c_{pqmr}, (p,q = 1,2,3)$ (characterized by Lam\'e parameters $\mu$, $\lambda$) within which there are randomly distributed string-like defects that can oscillate around their equilibrium position. The variables describing the system are the material displacements ${\mathbf u (\mathbf x,t)}$, at time $t$, of a point whose equilibrium position is ${\mathbf x}$. In addition, there are strings described by a vector ${\mathbf X} (s,t;{\bf x}_0,{\bf L},\sphericalangle)$, where $0<s<L$ is a position parameter along the string whose ends are fixed at ${\bf x}_0$ and ${\bf x}_0 + {\bf L}$. An orientation $\sphericalangle$ is needed in order to specify the glide plane (see Fig.~\ref{fig:dislocation-drawing}).

These strings are {Volterra} dislocations~\cite{Volterra1907}: the displacements ${\mathbf u (\mathbf x,t)}$ are multi-valued functions with a discontinuity equal to a Burgers vector ${\bf b}$ when crossing a surface, part of whose boundary includes the string with fixed ends. In addition to this geometrical fact, the coupling between elastic displacements and the elastic string is given by standard conservation of energy and momentum arguments. When dislocation velocities are small compared to the speed of sound, an assumption we shall make throughout this work, this leads to the well-known Peach-Koehler force \cite{Peach1950}.  In the time-dependent case, and for string velocities small compared to the speed of sound, the dynamics is described by the following classical action \cite{Lund1988}:
\begin{equation}
    S= S_{\rm ph} + S_{\rm string} + S_{\rm int} + S_0 \label{eq:actionFL}
\end{equation}
where 
\begin{align}
    S_{\rm ph} &= \frac{1}{2} \int \! dt \! \int \! d^3 x  \! \left( \rho \dot{\mathbf u}^2 - c_{pqmr} \frac{\partial u_m}{\partial x_q} \frac{\partial u_p}{\partial x_r} \right) + \cdots \, , \\
    S_{\rm string} &= \frac{1}{2} \int \! dt \! \int_{{\bf x}_0, L, \sphericalangle } \int_0^L \!\!\! ds  \! \left( m \dot{\mathbf X}^2 - \Gamma {\mathbf X}'^2 \right) + \cdots \, , \\
S_{\rm int} &= - b_i \int \! dt \! \int_{{\bf x}_0,  L, \sphericalangle } \int_{\delta \mathcal{S}} dS^j \sigma_{ij} \, ,
\end{align}
where $\sigma_{ij}=c_{ijkl}\partial u_k /\partial x_l$,  the ellipses ``$\cdots $'' refer to higher order terms in the phonon or string actions, and $S_0$ involves the interaction of the elastic displacements with a static dislocation, part of whose boundary is the straight line. $S_0$ does not contribute to the dynamics. The parameters $m$ and $\Gamma$ characterize the dislocation segments~\cite{Lund1988}. They may be written in terms of a dimensionless coupling constant $g$ and the Burgers vector ${\bf b}$ as
\begin{align}
    m &= \frac{\rho b^2}{g^2} \, , \\
    \Gamma &= \alpha^2 c_T^2 m \, ,
\end{align}
where $\alpha = \sqrt{2(1-\gamma^{-2})/(1+\gamma^{-4})}$, $\gamma = c_L/c_T$, $c_T = \sqrt{\mu/\rho}$ is the speed of propagation of shear waves, and $c_L$ that of pressure waves. $\alpha c_T$ is the velocity of propagation of perturbations on the dislocations. In terms of quantities used in previous works, the coupling strength may be written as $g = \sqrt{4\pi/\ell}$, with $\ell = (1+\gamma^{-4}) \ln (\delta /\delta_0)$ and $\delta$, $\delta_0$ microscopic cutoff lengths. Our attitude towards $g$ in this work will be to treat it as an adjustable parameter that characterizes the theory. The reason to call it a coupling constant is that upon rewriting the free string action using a ``canonical'' normalization such that $S_{\rm string} = \frac12 \int \dot{\bf X}^2 + \cdots$, the interaction term $S_{\rm int}$ becomes proportional to $g$. Thus, $g$ quantifies the strength of the phonon-dislocation interactions.

The quantization of this model proceeds as discussed in previous work of ours~\cite{Lund2019}. The results of that paper that are of highest relevance for our present work are contained in Section IV. C. thereof. The one crucial (yet straightforward) new ingredient we add is to consider a distribution of string lengths $p(L)$, which takes the place of the dislocation density $n_d$ by substituting it for $\int dL \, p(L)$ whenever it appears. We now discuss this additional ingredient in detail.

\subsection{The length distribution and the boson peak} \label{sec:pL-bosonpeak}

As we introduced earlier, the existence of the boson peak in amorphous solids quantifies the presence of additional vibrational degrees of freedom beyond those captured in the Debye model, and it has been shown that these can be satisfactorily described in terms of the vibrational degrees of freedom of linear defects, which we also refer to as dislocation segments or strings, in the material.

To connect the dislocation density per unit of length $p(L)$ with the boson peak density of states $g_S(\omega)$, we will use the same relation proposed by BGL\cite{Bianchi2020}, in which each ``string'' has only one vibrational mode, with which
\begin{equation}
    g_S(\omega_0) d\omega_0 = p(L) dL \, . \label{eq:BPdos-pL}
\end{equation}
BGL included a factor of $2$ on the r.h.s. to account for the two directions of oscillation. Here we will count each direction of oscillation as an additional string. In this expression, $g_S(\omega)$ corresponds to the volume density of vibrational degrees of freedom per unit of frequency in excess of that described by the Debye model, i.e., it accounts for the boson peak. $\omega_0$ is related to $L$ via the dispersion relation of the fundamental mode of the string, $\omega_0 = \alpha c_T \pi/L$.

In light of the string picture, it should be noted that each dislocation segment does not only have one vibrational mode: in the continuum limit, there exist an infinite number of modes characterized by wavenumbers $n\pi/L$, with $n \in \{1,2,3,\ldots\}$, all of which contribute to the observed boson peak in this description. The complete expression that takes the place of Eq.~\eqref{eq:BPdos-pL} is
\begin{equation}
    g_S(\omega_0) d\omega_0 = \sum_{n=1}^{\infty} p(nL) d(nL) \, , \label{eq:BPdos-pL-complete}
\end{equation}
where $L$ and $\omega_0$ are related as before. This expression indicates that the density of states at frequency $\omega_0$ receives contributions from the fundamental mode as above ($n = 1$), but also from the $n$-th excited states of strings with length $nL$.

However, the dislocation densities BGL extracted behaves as $L^{-5}$ at large $L$, meaning that, in practice, the modification of $p(L)$ relative to their fit that Eq.~\eqref{eq:BPdos-pL-complete} implies is a multiplicative factor of $\sum_{n=1}^\infty \tfrac{n}{n^5} = \zeta(4) \approx 1.08$, which is not a large modification. We will therefore continue using Eq~\eqref{eq:BPdos-pL} in our description, as it will not alter our qualitative conclusions.

We now turn to recalling the available data on the boson peak of the two materials we will study in detail later on: glycerol and silica.

\subsubsection{The boson peak and length distribution in Glycerol}

For the case of Glycerol, the BP DOS does not vary appreciably as a function of temperature below $T = 170$ K, as may be seen from data~\cite{Wuttke1995}. Because of this, we will simply use the single temperature result from BGL, which provided a rather good fit to the BP data across the entire frequency range (see Figure 1 therein~\cite{Bianchi2020}). We give the explicit form of this fit in Appendix~\ref{app:DOS-fits}. The only difference with their treatment in what follows is that, as stated before, we will use $\omega_0 = \alpha c_T \pi/L$ instead of setting $\alpha \to 1$, and we will not include the factor of $2$ when obtaining $p(L)$.

\begin{figure}
    \centering
    \includegraphics[width=0.95\linewidth]{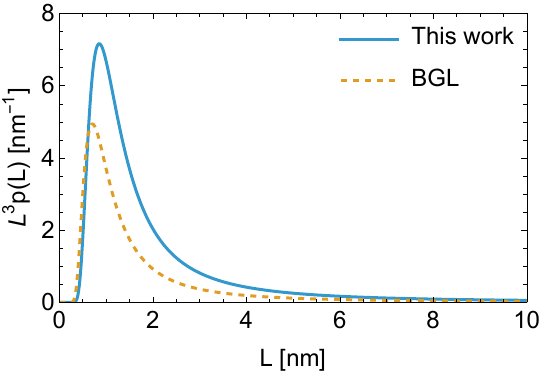}
    \caption{String length distribution in Glycerol determined from the BP density of states determined by BGL\cite{Bianchi2020} (see their Figure 1). For comparison, we display the length distribution BGL extracted from the same BP data, accounting for the factor of $2$ in the mode counting but maintaining the (mild) discrepancy between the dispersion relation of the string modes. For this figure, we choose $c_T$ to be equal to the physical speed of shear waves in Glycerol.}
    \label{fig:pL-glycerol}
\end{figure}

We present the resulting $p(L)$ in Figure~\ref{fig:pL-glycerol}. We can see that there is a mild shift in the position of the maximum of the distribution, as expected due to using a slightly different dispersion relation for the vibrational modes of the strings, but otherwise maintaining every feature of the distribution. %\footnote{
For this figure, we choose $c_T$ to be equal to the physical speed of shear waves in Glycerol. However, as will become clear later, the value of the velocity that enters the mapping between the BP DOS and the length distribution is the ``bare'' $c_T$, i.e., the quantity derived from the Lagrangian parameters $\mu$ and $\rho$. This means that, depending on the coupling strength between phonons and dislocations, which correspond to different bare velocities for a fixed physical velocity, the extracted distribution $p(L)$ will be quantitatively different. We show these changes in Appendix~\ref{app:DOS-fits}. %}

\subsubsection{The boson peak and length distribution in Silica}

For the case of Silica, the BP density of states has been measured across a wide range of temperatures\cite{}, and it exhibits a nontrivial temperature dependence.
Because of this, we will proceed to find the dislocation density as a function of both dislocation length $L$ and temperature $T$, obtaining $p(L,T)$. Given the available data~\cite{Wischnewski1998,Fontana1999}, we use a linear interpolation model between $T_1 = 51$ K and $T_2 = 1673$ K to fit the boson peak to an analytic function of $\omega_0$. That is to say, we take
\begin{equation}
    p(L,T) = \frac{T_2 - T}{T_2 - T_1} p(L, T_1) + \frac{T - T_1}{T_2 - T_1} p(L,T_2) \, . \label{eq:pL-interp}
\end{equation}

\begin{figure}
    \centering
    \includegraphics[width=0.95\linewidth]{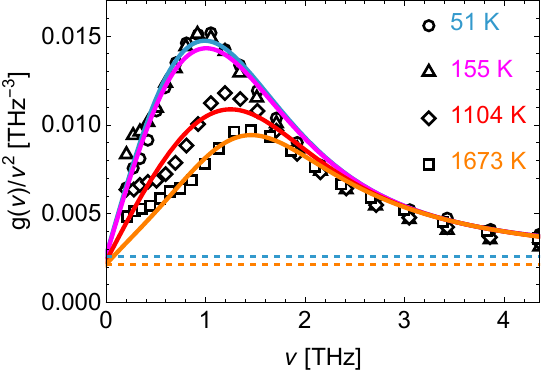}
    \caption{Data from Fig. 2 in Wischnewski et al.\cite{Wischnewski1998}, showing the boson peak data in amorphous SiO${}_2$ (open symbols), over which we display the parametrization we use to describe it in our model (solid colored lines). This parametrization is a linear (affine) function of the temperature. The BP density of states, which we use to obtain $p(L)$, is the difference between the data and the dashed horizontal lines that yield the Debye density of states at 51 K and 1673 K (drawn with their corresponding colors in the fit).
    }
    \label{fig:silica-DOS-fit}
\end{figure}

\begin{figure}
    \centering
    \includegraphics[width=0.95\linewidth]{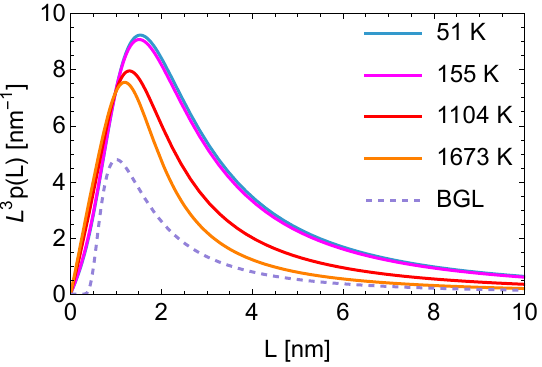}
    \caption{String length distribution in Silica determined from the BP density of states in Fig.~\ref{fig:silica-DOS-fit} (solid lines), for the same temperatures as shown in Fig.~\ref{fig:silica-DOS-fit}. The temperature dependence of the distribution is determined from Eq.~\eqref{eq:pL-interp} using the distribution at $T_1 = 51$ K and $T_2 = 1673$ K as input. For comparison, we display the length distribution BGL\cite{Bianchi2020} extracted from the $T = 1673$ K data using a different parametrization (dashed line).  
    }
    \label{fig:pL-silica}
\end{figure}

Following BGL, and guided by the data, we assume that the low-frequency behavior of $g_S(\omega)$ is proportional to $\omega^3$, to match the roughly linear trends seen in the data for $g_S(\omega)/\omega^2$. We present our fits in Figure~\ref{fig:silica-DOS-fit} compared with data\cite{}, and then the corresponding dislocation density curves $p(L,T)$ in Figure~\ref{fig:pL-silica}. We also compare with the results from BGL. We note that our fits yield a defect length distribution that is around a factor of 2 larger than that obtained by BGL. This is because the ``bare'' parameters we use as input in the Lagrangian $c_L$, $c_T$ are not matched directly to the physical speeds of sound propagation, but rather they are tuned so that the predictions for these speeds match data \textit{after} interactions are taken into account, i.e., after the calculation is complete. Their shape also differs slightly because we have not used the same functional form as they did at all frequencies. We give the explicit form of our fits in Appendix~\ref{app:DOS-fits}.

\subsection{The phonon propagator}

Let the Fourier modes of the displacement fields be defined as
\begin{equation}
    u_i({\bf x},t) = \int_{\bf k} \int_{-\infty}^{\infty} \frac{d\omega}{2\pi}  e^{-i\omega t} e^{i {\bf k} \cdot {\bf x}} u_i({\bf k}, \omega) \, ,
\end{equation}
where $\int_{\bf k} \equiv \int \frac{d^3k}{(2\pi)^3}$.

Given that the theory at hand is quadratic in its degrees of freedom, the statistics of the displacement fields are Gaussian, fully specified by two-point correlation functions. Assuming an isotropic and homogeneous medium, the phonon two-point correlator can be decomposed in terms of transverse and longitudinal components, i.e,
\begin{align}
    \langle u_i({\bf k}', \omega') & u_j({\bf k},\omega) \rangle \\ &= (2\pi)^4 \delta^3({\bf k} + {\bf k'}) \delta(\omega + \omega') \nonumber \\ &\times \left[ \left( \delta_{ij} - \hat{k}_i \hat{k}_j \right) \Delta^>_T({\bf k}, \omega ) + \hat{k}_i \hat{k}_j \Delta^>_L({\bf k}, \omega ) \right] \, , \nonumber
\end{align}
where $\Delta^>$ are called Wightman functions.

The main conceptual difference with our previous work\cite{Lund2019,Lund2020} is that in the setup we consider in this work these two-point functions are explicit functions of the temperature of the system. Hence, their calculation must be formulated at finite temperature from the beginning, in contrast to the calculation of scattering amplitudes of phonons by dislocations while the rest of the system is in its ground state. 

That being said, because the theory is quadratic, it turns out that the calculation of these two-point functions, including all of the effects of phonon-dislocation interactions, is operationally equivalent to that of the scattering amplitudes discussed in\cite{Lund2019} if the calculation is first carried out in Euclidean time, and then analytically continued to real time. We discuss this in the next section when we formulate the calculation of the thermal conductivity.

\section{Thermal conductivity via the Kubo formula}

The main ingredient of our calculation is the Kubo formula for the thermal conductivity, namely (setting $k_B = \hbar = 1$)
\begin{equation}
    \kappa_{ij} = \frac{1}{V T^2} \int_0^\infty dt \langle J_i(t) J_{j}(0) \rangle \, ,
\end{equation}
where $V$ is volume, $T$ is temperature, and $J_i$ is the spatially integrated energy current:
\begin{equation}
    J_i =  - c_{ijkl} \int_{\bf x} \frac{\partial u^j}{\partial t} \frac{\partial u^k}{\partial x^l} \, .
\end{equation}
We will use upper and lower indices interchangeably.
Note that the heat current $J_i$ only contains phonon fields, as they are the only degrees of freedom that propagate across space.

After a lengthy yet straightforward calculation, one finds that
\begin{align}
    \kappa_{ij} = \frac{\delta_{ij}}{3T^2} \int_{\bf k} & \int_{-\infty}^{\infty} \frac{d\omega}{2\pi} \omega^2 k^2 \nonumber \\ \Big[ & 2 \mu^2 \Delta_T^>({\bf k},\omega) \Delta_T^>(-{\bf k},-\omega) \nonumber \\ & + (\lambda + 2\mu)^2 \Delta_L^>({\bf k},\omega) \Delta_L^>(-{\bf k},-\omega) \nonumber \\ & + (\lambda + \mu)^2 \Delta_T^>({\bf k},\omega)  \Delta_L^>(-{\bf k},-\omega) \Big] \, .
\end{align}
As such, we can obtain the thermal conductivity directly from the phonon propagator $\Delta_{T/L}^>(\omega)$.

\subsection{Thermal conductivity in terms of phonon spectral functions}

Because of the thermal nature of the problem, the simplest way to proceed (conceptually) is to calculate the phonon propagator in \textit{Euclidean} time, i.e., to calculate
\begin{equation}
    \Delta^E({\bf x},\tau)_{ij} = \frac{1}{Z} {\rm Tr} \left[ u_i(\tau,{\bf x}) u_j(0,{\bf 0}) e^{-\beta H} \right] \, ,
\end{equation}
where Euclidean time evolution amounts to $\mathcal{O}(\tau) = e^{\tau H} \mathcal{O}(0) e^{-\tau H}$. The Hamiltonian $H$ is the Legendre transform of the action~\eqref{eq:actionFL}, and $Z = {\rm Tr} \left[ e^{-\beta H} \right]$ is the partition function of the theory.

As it turns out, the diagrammatic rules, the power series expansion, and everything else we worked through in real time in previous work\cite{Lund2019} (including the calculation of the self-energy), follows through identically, with the only difference that the Fourier modes of this correlator are labeled by their Matsubara frequencies $k_n = 2\pi n T $, $n \in \mathbb{Z}$ instead of the real-time frequency $\omega$.

Using the action in Eq.~\eqref{eq:actionFL} leads to an Euclidean propagator for the long-wavelength phonons given by
\begin{align}
    \Delta^E_T(k,k_n) &= \frac{\hbar/\rho}{ k_n^2 + c_T^2 k^2 [1 - \Sigma_T(k,k_n)] } \, , \\
    \Delta^E_L(k,k_n) &= \frac{\hbar/\rho}{ k_n^2 + c_L^2 k^2 [1 - \Sigma_L(k,k_n)] } \, ,
\end{align}
where
\begin{align}
    \Sigma_T(k,k_n) &= \int dL \, p(L) \frac{L \mu^2 b^2 }{4 m \rho c_T^2} {\rm Tr}_{n_D} \!\! \left[ \frac{F_T(k,k_n)}{1 - T(k_n)} \right] \, , \label{eq:selfenergyT-euclidean} \\
    \Sigma_L(k,k_n) &= \int dL \, p(L) \frac{L \mu^2 b^2 }{4 m \rho c_L^2} {\rm Tr}_{n_D} \!\! \left[ \frac{F_L(k,k_n)}{1 - T(k_n)} \right] \, . \label{eq:selfenergyL-euclidean}
\end{align}
where the trace ${\rm Tr}_{n_D}$ is defined in~\cite{Lund2019} and corresponds to a sum over the vibrational modes of the string.

Now we want to calculate $\Delta_T^>$ and $\Delta_L^>$. As it turns out, these correlation functions can be calculated using standard thermal field theory machinery~\cite{Laine:2016hma}. First, we note that this type of correlator is related to a so-called spectral function by
\begin{align}
    \Delta_T^>({\bf k},\omega) &= \frac{e^{\omega/T}}{e^{\omega/T} - 1} \rho_T({\bf k}, \omega) \, , \\
    \Delta_L^>({\bf k},\omega) &= \frac{e^{\omega/T}}{e^{\omega/T} - 1} \rho_L({\bf k}, \omega) \ ,
\end{align}
and these spectral functions can be obtained by analytically continuing the Euclidean correlator\cite{Laine:2016hma}:
\begin{align}
    \rho_T({\bf k}, \omega) &= 2 \, {\rm Im} \left\{ \Delta_T^E({\bf k}, k_n \to -i\omega + 0^+) \right\} \, , \\
    \rho_T({\bf k}, \omega) &= 2 \, {\rm Im} \left\{ \Delta_L^E({\bf k}, k_n \to -i\omega + 0^+) \right\} \, .
\end{align}
We have set $\hbar = 1$ in the above as well.

Because all functions involved are manifestly finite/analytic, the analytic continuation is direct. The only terms that deserve some care are the instances of $|k_n| = \sqrt{k_n^2}$, which is to be analytically continued to
\begin{equation}
    \sqrt{k_n^2} \to \sqrt{-\omega^2 - i {\rm sgn}(\omega) 0^+ } = - i \omega \, ,
\end{equation}
which also means $|k_n|^3 \to i \omega^3$.
Furthermore, since the analytically continued self-energies have a non-infinitesimal imaginary part, the role of the $0^+$ prescription is secondary in what follows.

If we take the imaginary parts of the previous expressions, we arrive at a transverse spectral function given by
\begin{align} \label{eq:rhoT-expr}
    &\rho_T(k,\omega) \\ &=  \frac{ (2/\rho)  c_T^2 k^2 {\rm Im}\{\Sigma_T\} }{\left(-\omega^2 + c_T^2 k^2 \left[ 1 -  {\rm Re}\{\Sigma_T\} \right] \right)^2 + \left(c_T^2 k^2 {\rm Im}\{\Sigma_T\} \right)^2 }    \, , \nonumber
\end{align}
and a longitudinal one
\begin{align} \label{eq:rhoL-expr}
    &\rho_L(k,\omega) \\ &=  \frac{ (2/\rho) c_L^2 k^2 {\rm Im}\{\Sigma_L\} }{\left(-\omega^2 + c_L^2 k^2 \left[ 1 -  {\rm Re}\{\Sigma_L\} \right] \right)^2 + \left(c_L^2 k^2 {\rm Im}\{\Sigma_L\} \right)^2 } \, , \nonumber
\end{align}
and the thermal conductivity is then given by
\begin{align}
    \kappa_{ij} = & \frac{\delta_{ij}}{12 \pi^3 T^2}  \int_0^{k_{\rm cut}} \!\!\!\!\! dk \int_{-\infty}^{\infty} \!\!\!  d\omega \frac{ \omega^2 k^4 e^{\omega/T}}{(e^{\omega/T}-1)^2}  \label{eq:conductivity} \\ 
    & \times \left[ 2\mu^2 |\rho_T|^2 + (\lambda + 2\mu)^2 |\rho_L|^2 + (\lambda + \mu)^2 |\rho_T \rho_L| \right] \, , \nonumber
\end{align}
where $k_{\rm cut}$ is a microscopic wavenumber cutoff indicating the breakdown of the continuum theory we employ here. We will set this momentum space cutoff to be $k_{\rm cut} = \pi/ a$, where $a$ is the typical intermolecular distance in the glass.
This is the starting point for the rest of our analysis.

\subsection{Low temperature behavior in the string model}

It follows from the previous considerations that, in the string model specified by the action principle~\eqref{eq:actionFL}, the contribution to the phonon self-energies from scattering by dislocations is given by~\cite{}
\begin{align}
    \Sigma_T(\omega) &= \frac{\omega}{c_T} \frac{g^2}{4} \int_0^\infty \frac{dx}{x^2} \, G \! \left( \frac{x c_T}{\omega} \right) {\rm Tr}_{n_D} \!\! \left[ \frac{\tilde{F}_T(x) }{ D(x) - \tilde{T}(x) } \right]    \, , \label{eq:SigmaT} \\
    \Sigma_L(\omega) &= \frac{\omega}{c_T} \frac{g^2}{4 \gamma^2} \int_0^\infty \frac{dx}{x^2} \, G \! \left( \frac{x c_T}{\omega} \right) {\rm Tr}_{n_D} \!\! \left[ \frac{\tilde{F}_L(x) }{ D(x) - \tilde{T}(x) } \right]  \, , \label{eq:SigmaL}
\end{align}
where, to most clearly exhibit the properties of these expressions, we have rescaled the frequency $\omega$ out of all functions except from
\begin{equation}
    G(L) = p(L) L^5 \, ,
\end{equation}
and the rest of the functions ($\tilde{F}_{T/L}$, $D$, $\tilde{T}$) are defined in Appendix~\ref{app:aux-expr}, where we also explain the aforementioned rescaling. Throughout our work in what follows, we have made one further simplification relative to Eqs.~\eqref{eq:selfenergyT-euclidean} and~\eqref{eq:selfenergyL-euclidean} in that we have replaced $\Sigma(k,\omega) \to \Sigma(\omega) = \Sigma(k(\omega),\omega) $, with $k(\omega)$ the free phonon dispersion relation. This is an approximation motivated by the fact that we expect that, while impeded by scattering with line defects, most of the contribution to the heat flux will come from phonons that are not far from being on-shell. In Appendix~\ref{app:Ioffe-Regel} we quantify, albeit indirectly, the degree to which this is an accurate approximation by examining whether the Ioffe-Regel limit is reached in our setup, after all other calculations have been carried out.  %\fl{BSH: justificar? R: describ\'i un poco m\'as.}

It would seem from here that if $p(L) \propto L^{-5}$ at large $L$, then the self-energies are linear functions of the frequency $\omega$ at low frequencies. This is so because all of the nontrivial (in this case, nonlinear) dependence of $\Sigma_{T/L}$ on $\omega$ appears through $G$, where the low frequency limit corresponds to the large $L$ limit, and the scenario $p(L) \propto L^{-5}$ yields $G \propto L^0$, meaning that no additional $\omega$ dependence (beyond the linear prefactor) would appear in this limit. 
As it turns out, this is only partially true: if $G(L)$ is exactly constant, then the integrand is actually divergent at small $x$. Therefore, the actual value of the integral does depend on the $L$ dependence of $p(L)$ at small $L$, regardless of how small $\omega$ may be. As a consequence of this, the real part of the self-energy contains a contribution proportional to $\omega^0$ (constant) at small $\omega$, which is dominated by the distribution of short length linear defects in the material. 
The net effect of this contribution is to renormalize the speed of propagation of phonons. We denote these renormalized speeds by $\tilde{c}_{T/L}$.
On the other hand, the imaginary part of the self-energy receives no divergent contribution at small $x$, and so it is truly linear in frequency at small $\omega$.

This dependence enables us to predict how the thermal conductivity scales with temperature in such a material at low temperatures. Starting from Eq.~\eqref{eq:conductivity}, taking $T \to 0$ means that only low frequency contributions will be significant, due to the fact that $\omega^2 e^{\omega/T}/(e^{\omega/T}-1)^2$ becomes more and more localized around $\omega = 0$ in this limit. Then, the spectral functions $\rho_T$ and $\rho_L$ become
\begin{align}
    \rho_T(k,\omega) &\approx \frac{2}{\rho} \frac{A_T c_T k^2 \omega}{(-\omega^2 + \tilde{c}_T^2 k^2)^2 + A_T^2 c_T^2 k^4 \omega^2 } \, , \\
    \rho_L(k,\omega) &\approx \frac{2}{\rho} \frac{A_L c_L k^2 \omega}{(-\omega^2 + \tilde{c}_L^2 k^2)^2 + A_L^2 c_L^2 k^4 \omega^2 } \, ,
\end{align}
where $A_T, A_L$ are constants that characterize the magnitude of the (linear) low frequency dependence of ${\rm Im}\{ \Sigma_T \}$. Explicitly, they are determined by
\begin{align}
    A_T &= \frac{g^2 G_\infty}{4} \int_0^\infty \frac{dx}{x^2}\, {\rm Im} \! \left\{ {\rm Tr}_{n_D} \!\! \left[ \frac{\tilde{F}_T(x) }{ D(x) - \tilde{T}(x) } \right] \right\} \, , \label{eq:AT-matching} \\
    A_L &= \frac{g^2 G_\infty}{4 \gamma} \int_0^\infty \frac{dx}{x^2}\, {\rm Im} \! \left\{ {\rm Tr}_{n_D} \!\! \left[ \frac{\tilde{F}_L(x) }{ D(x) - \tilde{T}(x) } \right] \right\} \, , \label{eq:AL-matching}
\end{align}
Here we have introduced $G_{\rm \infty} = \lim_{L\to \infty} G(L) = \lim_{L \to \infty} p(L) L^5 $.

It is then apparent that the dominant contribution in the limit $T \to 0$ will be from the integration domain near the lines $\omega^2 = \tilde{c}_{T/L}^2 k^2$. Concretely, in this regime we have
\begin{align}
    \rho_T^2 &\to \frac{2\pi}{\rho^2 A_T c_T k^2 |\omega|} \delta(\omega^2 - \tilde{c}_T^2 k^2 ) \, , \\ 
    \rho_L^2 &\to \frac{2\pi}{\rho^2 A_L c_L k^2 |\omega|} \delta(\omega^2 - \tilde{c}_L^2 k^2 ) \, , \\ 
    \rho_T \rho_L &\to 0 \, .
\end{align}
    
Using these expressions in Eq.~\eqref{eq:conductivity}, one readily arrives at
\begin{align}
    \kappa &= \frac{1}{12\pi^2 T^2} \left[  \frac{c_T^3}{A_T \tilde{c}_T^3 } + \frac{c_L^3}{2 A_L \tilde{c}_L^3 } \right] \int_0^\infty \frac{d\omega \,  \omega^2}{\sinh^2 (\omega/2T) } \nonumber \\ 
    &=\frac{T}{9} \left[  \frac{c_T^3}{A_T \tilde{c}_T^3 } + \frac{c_L^3}{2 A_L \tilde{c}_L^3 } \right] \, , \label{eq:kappa-lowT}
\end{align}
with $\kappa_{ij}=\kappa \delta_{ij}$, thus demonstrating the linear rise in thermal conductivity at low (but above the plateau)  temperatures for materials with a density of linear defects $p(L) \propto L^{-5}$, as described above.

\subsubsection{What if the dislocation density had a different power law tail?}

It is pertinent to ask to what extent our results for the thermal conductivity depend on the length distribution falling off as $L^{-5}$ at large $L$. Indeed, at very large $L$, the length distribution is extracted from the very low frequency limit of the BP DOS according to Eq.~\eqref{eq:BPdos-pL} or Eq.~\eqref{eq:BPdos-pL-complete}, and an $\omega^3$ power law is not the only possibility that gives a reasonable description of data. For example, other authors~\cite{Moriel2024} have argued for describing the BP DOS at low frequencies as being proportional to $\omega^4$, or $a_2 \omega^2 + a_4 \omega^4$. 

Let us therefore assume that $g_S(\omega_0) \propto \omega_0^{\xi+3}$ at small $\omega_0$, or equivalently, $p(L) \propto L^{-(\xi+5)}$ at large $L$. It is then appropriate to define
\begin{equation}
    G_\xi(L) = p(L) L^{\xi + 5} \, ,
\end{equation}
with which the phonon self-energies become
\begin{align}
    &\Sigma_T(\omega) = \\ & \frac{\omega^{1+\xi}}{c_T^{1+\xi}} \frac{g^2}{4} \int_0^\infty \frac{dx}{x^{2+\xi}} \, G_\xi \! \left( \frac{x c_T}{\omega} \right) {\rm Tr}_{n_D} \!\! \left[ \frac{\tilde{F}_T(x) }{ D(x) - \tilde{T}(x) } \right]    \, , \nonumber \\
    &\Sigma_L(\omega) = \\ & \frac{\omega^{1+\xi}}{c_T^{1+\xi}} \frac{g^2}{4 \gamma^2} \int_0^\infty \frac{dx}{x^{2+\xi}} \, G_\xi \! \left( \frac{x c_T}{\omega} \right) {\rm Tr}_{n_D} \!\! \left[ \frac{\tilde{F}_L(x) }{ D(x) - \tilde{T}(x) } \right]  \, . \nonumber
\end{align}

The same considerations we gave earlier apply \textit{mutatis mutandis} -- in particular, with redefined versions of $A_{T/L}$, which we denote by $A_{\xi T/L}$, which give the coefficients in front of the leading contributions to the imaginary part of the self-energy, proportional to $\omega^{1+\xi}$. Assuming this behavior holds for all of the relevant frequency range for the temperature of interest, one arrives at
\begin{align}
    \kappa &= \frac{1}{12\pi^2 T^2} \left[  \frac{c_T^3}{A_{\xi T} \tilde{c}_T^3 } + \frac{c_L^3}{2 A_{\xi L} \tilde{c}_L^3 } \right] \int_0^\infty \frac{d\omega \,  \omega^{2-\xi}}{\sinh^2 (\omega/2T) } \nonumber \\ 
    &= T^{1-\xi} \frac{ \Gamma(3-\xi) \zeta(2-\xi) }{3\pi^2} \left[  \frac{c_T^3}{A_{\xi T} \tilde{c}_T^3 } + \frac{c_L^3}{2 A_{\xi L} \tilde{c}_L^3 } \right] \, . \label{eq:kappa-lowT-mod}
\end{align}
Note that the integral that defines the conductivity is IR divergent if $\xi \geq 1$. That is to say, for there to be resistance to heat transport by long wavelength phonons, the density of dislocations has to fall off more slowly than $L^{-6}$. Fewer long dislocations would not sufficiently impede phonon propagation to generate a finite thermal conductivity.

It is interesting nonetheless to consider the borderline case $\xi = 1$, corresponding to a BP DOS proportional to $\omega^4$ at low frequencies. This will lead to an almost temperature-independent\footnote{Strictly speaking, the dependence would be logarithmic.} thermal conductivity controlled by the system size, which defines the IR cutoff of the integral. This indicates that, if the impact of the BP on the thermal conductivity of glasses is to be understood via the elastic string description we present in this work, the dominant effects have to come from the frequency interval in which the BP DOS grows approximately as $\omega^3$ even if its precise functional dependence may be different. It is possible that pursuing this line of thought can provide a phonon-plus-strings understanding of the plateau in the thermal conductivity in the temperature interval 1-20 K mentioned in the Introduction. Although interesting, we shall not consider this possibility here, as it would take us outside the scope of the present work.

We stress that from our point of view, regardless of the precise power law dependence, the data is quantitatively well described by a $\omega^3$ power law all the way between $\omega = 0$ and the maximum of $g_S(\omega_0)$. Therefore, we expect that while distortions to our predicted linear $T$ dependence of the thermal conductivity $\kappa$ could happen at very low temperatures when $\kappa$ is only sensitive to extremely low frequency phonon modes, our predictions should remain robust at temperatures above 30 K.\footnote{For example, looking at Figure~\ref{fig:silica-DOS-fit}, the BP DOS for Silica is well described by a power law $\omega^3$ for frequencies around and above 0.5 THz. Converting this frequency to a temperature, one obtains $T =  (\pi \times 10^{12} \, {\rm s}^{-1} ) \hbar/k_B \approx 24 \, {\rm K} $.}

\subsection{The ballistic limit}

Before moving on to comparisons with data, in order to validate our intermediate results and conclusions, it is helpful, whenever possible, to compare with other theoretical approaches to thermal conductivity calculations. 

In what follows we show how the thermal conductivity formula one would derive in a kinetic theory of phonons emerges from Eq.~\eqref{eq:conductivity}. The appropriate approximation to make in order to obtain this formula is to take phonons to have a well-defined dispersion relation $\omega = c_{T/L}k$ and treat the effects of scattering by defects perturbatively. This is the \textit{ballistic} limit of phonon propagation, where phonons (approximately) travel in straight lines as free particles and only rarely scatter off impurities -- which also makes clear why, in this limit, kinetic theory is the right framework to compare with. 

Mathematically, in our framework this corresponds to taking the limit where the self-energies $\Sigma_{T/L}$ are small, and the spectral densities may be evaluated as
\begin{align}
    \rho_T^2 &\to \frac{2\pi}{\rho^2 c_T^2 k^2 | {\rm Im}\{\Sigma_T\} | } \delta(\omega^2 - \tilde{c}_T^2 k^2 ) \, , \\ 
    \rho_L^2 &\to \frac{2\pi}{\rho^2 c_L^2 k^2 | {\rm Im}\{\Sigma_L\} | } \delta(\omega^2 - \tilde{c}_L^2 k^2 ) \, , \\ 
    \rho_T \rho_L &\to 0 \, .
\end{align}

It then follows that the thermal conductivity takes the form
\begin{align}
    \kappa_{\rm ballistic} &= \frac{1}{12\pi^2 T^2}  \int_0^\infty \!\!\! \frac{d\omega \,  \omega^3}{\sinh^2 (\omega/2T) } \label{eq:kappa-ballistic-general} \\ & \quad \times \left[  \frac{c_T^2 }{ \tilde{c}_T^3 {\rm Im}\{\Sigma_T(\omega)\}  } + \frac{c_L^2 }{2  \tilde{c}_L^3 {\rm Im}\{\Sigma_L(\omega) \} } \right] \, , \nonumber 
\end{align}
which is to be compared with the formula one obtains from kinetic theory (in terms of the heat capacity per mode $C(\omega) = \tfrac{\omega^2}{T^2}  \tfrac{e^{\omega/T} }{(e^{\omega/T}  - 1 )^2} $, with units where $\hbar = k_B = 1$)
\begin{align}
	\kappa &= \frac{1}{6\pi^2} \int_0^\infty \!\!\! d\omega \, \omega^2 \left[ 2c_T^{-1}  \tau_T(\omega) + c_L^{-1} \tau_L(\omega) \right] C(\omega) \nonumber \\
	&=  \frac{1}{12 \pi^2 T^2} \int_0^\infty \!\!\! \frac{ d\omega \, \omega^4}{ \sinh^2 (\omega/2T) } \left[ \frac{ \tau_T(\omega) }{ c_T  } + \frac{ \tau_L(\omega) }{ 2 c_L  }  \right] \, .
\end{align}
(The upper limit of integration should be replaced with a cutoff when appropriate.)

Therefore, in the limit where interactions are very weak, where we may take $\tilde{c}_{T/L} = c_{T/L}$, our expression in Eq.~\eqref{eq:kappa-ballistic-general} recovers the kinetic theory result provided that we identify
\begin{align}
	\tau^{-1}_T(\omega) &= \omega \, {\rm Im}\{\Sigma_T(\omega)\} \, , \label{eq:tauT-rel-optical-th} \\
	\tau^{-1}_L(\omega) &= \omega \, {\rm Im}\{\Sigma_L(\omega)\} \, . \label{eq:tauL-rel-optical-th}
\end{align}
This is actually a cross-check of our derivation, because these last two lines are a direct consequence of the optical theorem in quantum field theory~\cite{}.

In particular, if one wanted to reproduce our results in the low frequency limit we discussed in the previous subsection from a purely kinetic theory point of view, all one would need to do is prescribe that the contribution of scattering by linear defects has inverse relaxation times $\tau^{-1}_{T/L}(\omega)$ that grow as $\omega^2$ (or $\omega^{2+\xi}$ in the general case where the length distribution of defects decays as $p(L) \propto L^{-(\xi+5)}$ at large $L$).

\section{Fitting the data}

Equipped with Eq.~\eqref{eq:conductivity} and expressions for the phonon self-energies in Eqs.~\eqref{eq:SigmaT} and~\eqref{eq:SigmaL}, we are now in a position to assess the validity of the string model quantitatively, by comparing its predictions to experimental data.

\subsection{Approach 1: Ballistic transport limit}

The low temperature results we discussed in the last Section suggest a very simple approach to describing the thermal conductivity: treat phonons as particles propagating ballistically with a well-defined dispersion relation $\omega = \tilde{c}_{T/L} k$. To justify this away from the low-temperature regime, one needs to assume that $A_T, A_L$ are small. Physically, this means that the scattering is not strong enough to significantly alter how phonons propagate.

The only additional ingredient that is needed in order to have a qualitative model of the thermal conductivity is to introduce a cutoff in the sum over modes. For simplicity, we introduce a single, common frequency cutoff $\Lambda$ for both transverse and longitudinal modes. The frequency integral is the same as in Eq.~\eqref{eq:kappa-lowT} except for the upper limit of integration, and can be carried out analytically.

The result can be written succinctly as
\begin{equation}
    \kappa = \frac{k_B^2}{9\hbar} \frac{\Lambda}{ A_{\rm eff} }  \tilde{\kappa}\!\left(\frac{T}{\Lambda}\right) \, , \label{eq:cond-simple}
\end{equation}
where $\Lambda$ and $A_{\rm eff}$ are to be fit by comparison with experimental data, and we have restored units with explicit factors of $\hbar$ and $k_B$, and $\tilde{\kappa}(x)$ is a dimensionless quantity defined as
\begin{align}
    \tilde{\kappa}(x) \equiv x \bigg[ 1 + \frac{6}{\pi^2} \bigg( & \frac{1}{4x^2} \left( 1 - \coth \frac{1}{2x} \right) \nonumber \\ & + \frac{1}{x} \ln (1 - e^{-1/x} ) - {\rm Li}_2(e^{-1/x}) \bigg) \bigg] \, .
\end{align}
The fit constant $A_{\rm eff}$ can be related to the parameters in our previous discussion via $A_{\rm eff} = \left[  \tfrac{c_T^3}{A_T \tilde{c}_T^3 } + \tfrac{c_L^3}{2 A_L \tilde{c}_L^3 } \right]^{-1}$.

\begin{figure}[t]
    \centering
    \includegraphics[width=0.95\linewidth]{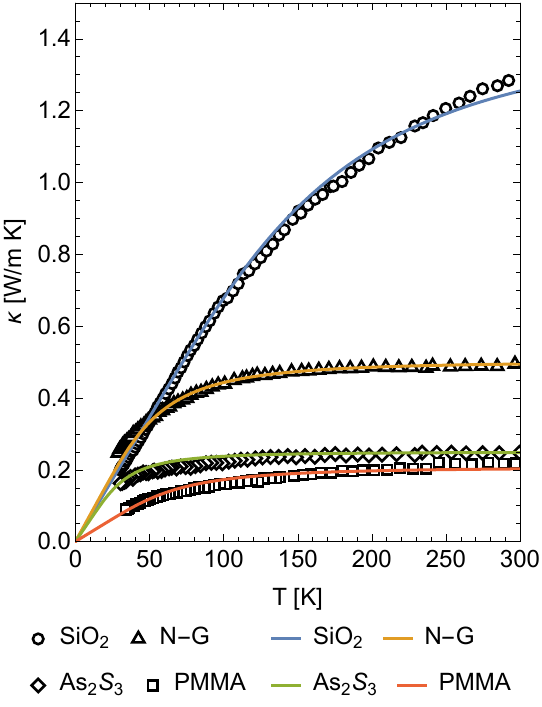}
    \caption{Comparison of data in Cahill et al.~\cite{Cahill1987}, reproduced from Figure 5 of that paper, with the model defined by Eq.~\eqref{eq:cond-simple} overlaid on top. The $y$-axis scale for SiO${}_2$ is on the right, whereas for all the others it is on the left. The values of $\Lambda$ and $A_{\rm eff}$ used for each model curve are presented in Table~\ref{tab:fit-params}.}
    \label{fig:conductivity-simple}
\end{figure}

\begin{table}[t]
    \centering
    \begin{tabular}{c|c|c}
        & $\Lambda$ [K] & $A_{\rm eff}$ [pm]  \\
         SiO${}_2$ & 680 & 29.01 \\
         N-G & 220 & 26.78 \\
         As${}_2$S${}_3$ & 135 & 33.06 \\
         PMMA & 270 & 79.74
    \end{tabular}
    \caption{Fit parameters used in Figure~\ref{fig:conductivity-simple}. Note that the units for $A_{\rm eff}$ are picometers $= 10^{-3}$ nanometers. The values of the frequency cut-off $\Lambda$ are not surprising in a condensed matter context. The cut-off values for $A_{\rm eff}$ can be rationalized using dimensional analysis as follows: In order to have a thermal conductivity that is a function of Boltzmann's constant and of Planck's constant and that also is a linear function of temperature $T$, it must have the form $\kappa = {\rm (const)} k_B^2 T/\hbar$ with (const) a constant with dimensions of inverse distance. Fitting this expression to the low-temperature end of the data in Figure \ref{fig:conductivity-simple} provides the values of $A_{\rm eff}$ above. }
    \label{tab:fit-params}
\end{table}

As we can see from Figure~\ref{fig:conductivity-simple}, this very simple model can quantitatively describe the thermal conductivity of glassy materials over a broad range of temperatures. Furthermore, the fit parameters have straightforward physical interpretations in terms of a ``Debye'' cutoff for $\Lambda$ and line defect properties $A_{\rm eff}$ (encoding a product of their density and interaction strength with phonons). Therefore, the picture that emerges is very satisfactory, as it provides insight into the microscopic physics of glasses.

So far, this result is a resounding success. However, we have not used all of the experimental data at hand to constrain our theory. Earlier on in Section~\ref{sec:pL-bosonpeak}, we argued that the density of dislocations $p(L)$ can be reconstructed from the boson peak of glasses. Therefore, we should be able to use all of this information to gain further insight into the microscopic physics at play, allowing us, for example, to determine the interaction strength between phonons and line defects, as well as to obtain additional cross-checks of the conclusions we have obtained thus far. This is what we do next.

\subsection{Glycerol with detailed DOS input: a successful description only with line defects}
\label{gly}

While the picture we just presented is quite satisfactory and can be described in terms of fairly simple expressions, our starting point contains much more information. Given full knowledge of $p(L)$, which we have in the case of SiO${}_2$ and Glycerol, we can use Eq.~\eqref{eq:conductivity} to do a complete calculation in the microscopic theory defined by Eq.~\eqref{eq:actionFL}, and determine the interaction strength $g$ between phonons and line defects in each material. The challenge is then to evaluate the integrals that define the phonon self-energies $\Sigma_{T/L}(\omega)$, and use them to obtain the spectral functions that enter the thermal conductivity integral.

In the case of Glycerol, we have somewhat less information than for SiO${}_2$ because in the latter we have data for a much broader temperature range for both the thermal conductivity and the speeds of longitudinal and transverse waves, and they exhibit a clear temperature dependence. As we will see later, in SiO${}_2$ it is possible to use the temperature dependence of the speed of sound to fix the value of $g$. Here we will treat the coupling constant $g$ as a free adjustable parameter because the lack of significant temperature dependence in the data we have available for the BP DOS and the speed of sound in Glycerol means we only have two numbers as data to constrain our model.

In any case, the value of $g$ must be consistent with the measured value of the speed of sound.
As we remarked upon earlier, the physical speed of sound $\tilde{c}_{T/L}$ is not the ``bare'' speed of sound $c_{T/L}$ that is directly related to the parameters in the Lagrangian as $c_T = \sqrt{\mu/\rho}$, $c_L = \sqrt{(\lambda+2\mu)/\rho}$, but rather a renormalized version thereof
\begin{align}
    \tilde{c}_T^2 &= c_T^2 \left[ 1 - \Sigma_T(\omega = 0) \right] \nonumber
    \\ 
    &=  c_T^2 \left[ 1 - \frac{g^2}{60 \alpha^2} \int_0^\infty dL \, L^3 p(L)  \right] \, , \label{eq:cT-phys} \\
    \tilde{c}_L^2 &= c_L^2 \left[ 1 - \Sigma_L(\omega = 0) \right]  \nonumber
    \\ 
    &=  c_L^2 \left[ 1 - \frac{g^2}{45 \alpha^2 \gamma^2} \int_0^\infty dL \, L^3 p(L)   \right] \, , \label{eq:cL-phys}
\end{align}
where, as before, $\gamma = c_L/c_T$, and we have carried out the sums over the vibrational degrees of freedom of the string analytically. Just like $g$, the two bare speeds of sound $c_L$, $c_T$ may be treated as adjustable parameters, with their values only being constrained by having an overall agreement with data after including the effects of interactions. With the data available to us, this fit is in fact under-constrained because we have three parameters and only two numbers to fit them.

While a unit cell does not exist in a disordered material, a microscopic length $a$ associated to the UV cutoff of the continuum description $k_{\rm cut} = \pi/a$ is still a necessary input to define the theory.
We have found that it is possible to arrive at a reasonable description of data if one sets the microscopic cutoff $a_{\rm Glycerol}$ to be around 0.7 -- 0.8 nm, and adjusting the coupling constant for each individual scenario. Whether a fit is better or worse depends on how one weighs the data, and so we do not present our results in terms of a ``best fit,'' but rather show the resulting curves for different possible scenarios.

\begin{figure}
    \centering
    \includegraphics[width=0.95\linewidth]{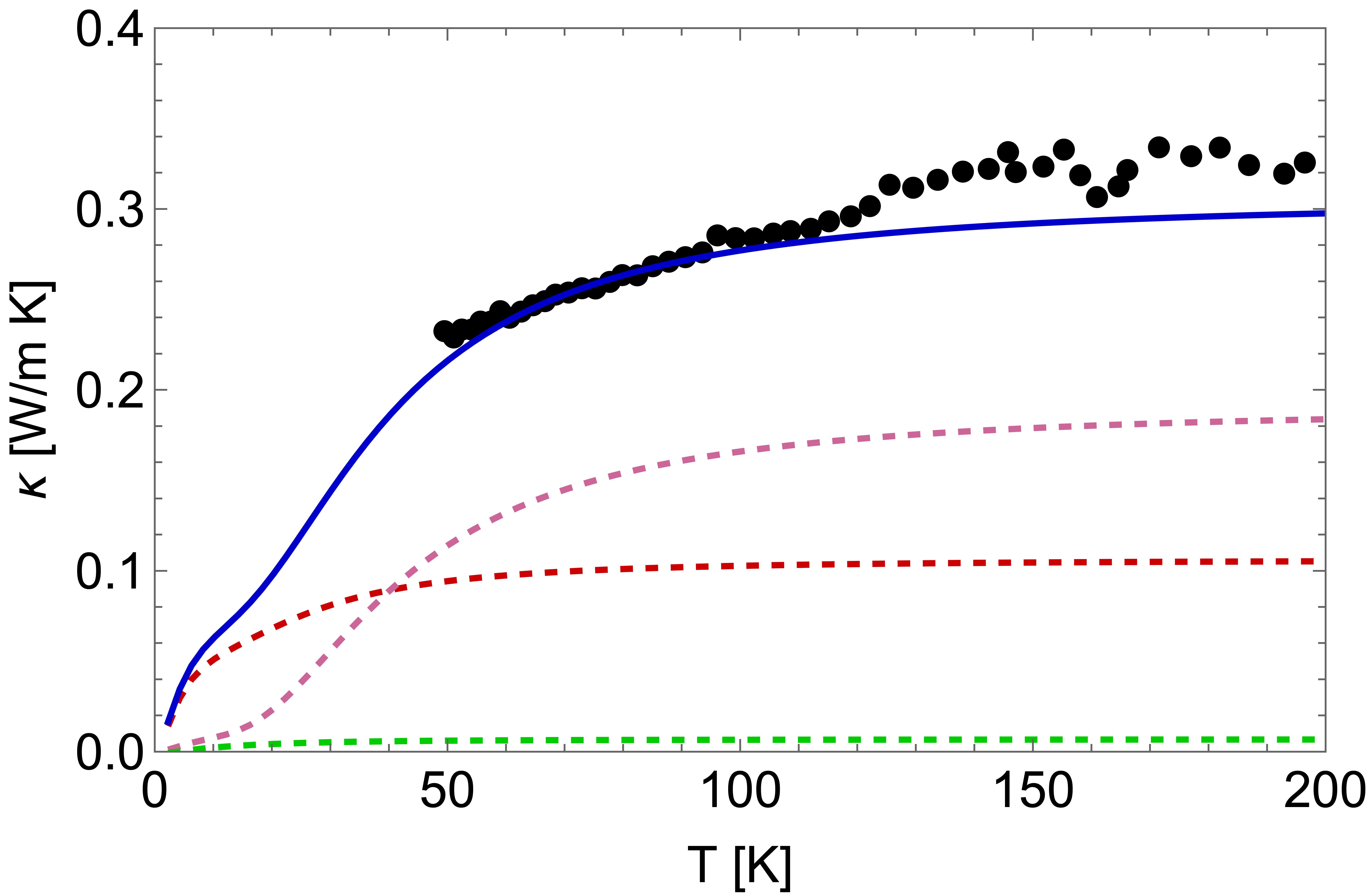}
    \includegraphics[width=0.95\linewidth]{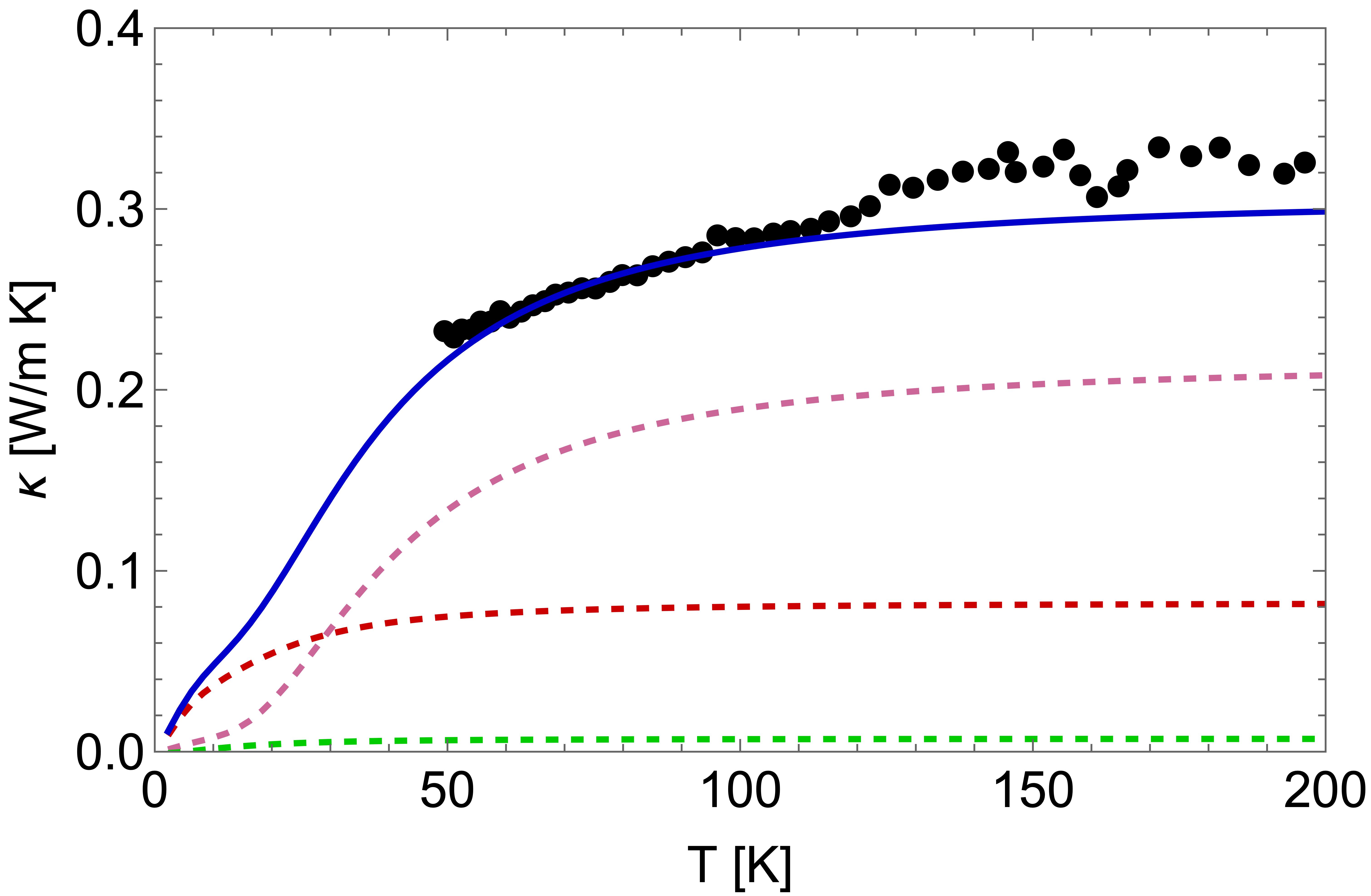}
    \includegraphics[width=0.95\linewidth]{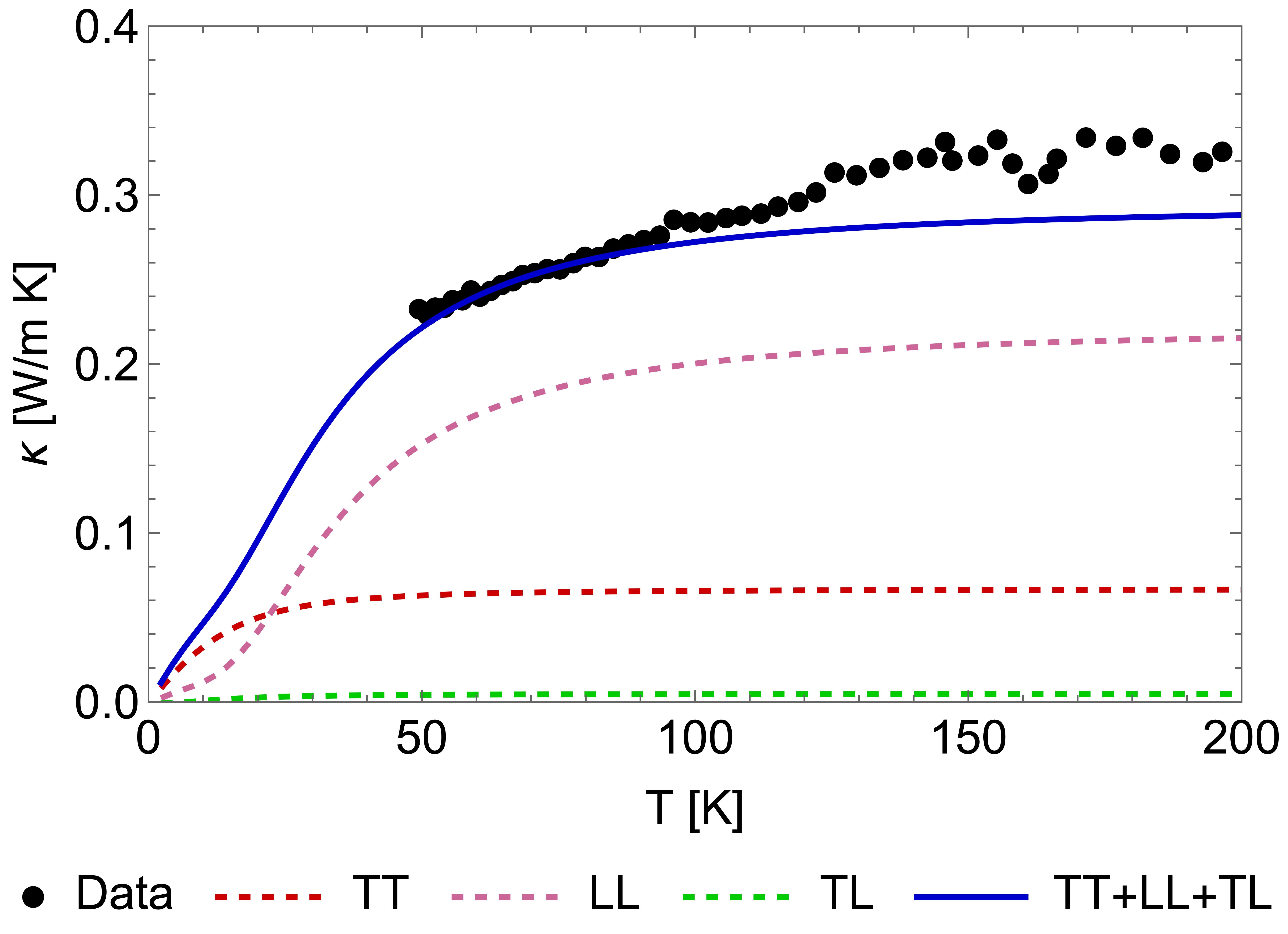}
    \caption{Selected fits for Glycerol thermal conductivity data for different values of the microscopic length cutoff $a_{\rm Glycerol}$ and coupling constant $g_{\rm Glycerol}$. In each case, we perform a fit to $(c_L,c_T)$ so that phonons propagate with the physical speed of sound $\tilde{c}_L = 3420$ m/s and $\tilde{c}_T = 1657$ m/s. Top: $a_{\rm Glycerol} = 0.77$ nm, $g_{\rm Glycerol} = 0.994$. Middle: $a_{\rm Glycerol} = 0.694$ nm, $g_{\rm Glycerol} = 1.204$. Bottom: $a_{\rm Glycerol} = 0.7$ nm, $g_{\rm Glycerol} = 1.315$.}
    \label{fig:Glycerol-cond}
\end{figure}

In Figure~\ref{fig:Glycerol-cond} we show results for three possible values of the coupling strength, given in each case by $g_{\rm Glycerol} = 0.994$, $g_{\rm Glycerol} = 1.204$, and $g_{\rm Glycerol} = 1.315$. The values of the microscopic length cutoff are $a_{\rm Glycerol} = 0.77$ nm, $a_{\rm Glycerol} = 0.694$ nm, and $a_{\rm Glycerol} = 0.7$ nm, respectively. The corresponding values of the bare speeds of propagation are ($c_L = 4863.6$ m/s, $c_T = 3422.7$ m/s), ($c_L = 4235.6$ m/s, $c_T = 2725.6$ m/s), ($c_L = 3799.6$ m/s, $c_T = 2191.3$ m/s). Comparing to the physical speeds of propagation $\tilde{c}_L = 3420$ m/s and $\tilde{c}_T = 1657$ m/s, we see that the modifications due to interactions with defects can be sizeable --- and in fact, that they must be so in order to describe both the physical speeds of propagation and the thermal conductivity. While none of the three are perfect fits to the data, they all give a reasonable description of the data in the range between 50 and 200 K. This is quite satisfactory given that, except for the value of the coupling constant and the microscopic length cutoff, all of the quantities in the theory are fixed. Note that while the sum over all conductivity channels (transverse TT, longitudinal LL and interference TL as defined, in an obvious notation, by Eq. \eqref{eq:conductivity}) is very similar, the individual contributions exhibit a significant variation in the three scenarios. 

Most importantly, all of said quantities are defined in terms of a continuum theory with no detailed treatment of the constituents of the material. By pursuing a microscopic description at the atomic level, one would not need to specify the microscopic length $a$ as an additional input, and $g_{\rm Glycerol}$ (or the corresponding parameter in such a description) would thus be uniquely determined. On the other hand, the parameters of a theory in the continuum can certainly be functions of the microscopic cutoff of the theory. Such dependence encodes the microscopic information that has been ``integrated out'' of the description, consistent with renormalization group considerations\cite{}. While it would be interesting to explore how the phonon-dislocation interactions get renormalized as the microscopic cutoff is varied, this is outside the scope of this work.

\subsection{Silica with detailed DOS and speed of sound input: a description with line and point defects}

Compared to Glycerol, glassy silica presents additional challenges. On the one hand, the BP in silica exhibits a quantitatively substantial temperature dependence (see Fig.~\ref{fig:silica-DOS-fit}). In addition to this, even though we have not discussed it thus far, the transverse and longitudinal speeds of sound in silica also display an unambiguous temperature dependence~\cite{}. While the variation in the speed of sound itself is a 10\% effect, because the conductivity may depend on $\tilde{c}_{T,L}$ through powers thereof, the ensuing variation could conceivably be of 30\% (e.g., see Eq.~\eqref{eq:kappa-ballistic-general}). Therefore, we have to make sure that our results are consistent with speed of sound data. 

As we have seen, a description of thermal conductivity in terms of phonons in interactions with dislons provides a satisfactory agreement with experimental data for SiO$_2$ up to 300 K, Ca-K nitrate, As$_2$O$_3$, PMMA and Glycerol. In the case of silica, where data is available for a much broader temperature interval, this picture, as we see below, while providing a reasonable qualitative explanation for the data trend, does not fit it in quantitative detail. Could there be an additional mechanism that is an obstacle to the energy transport by phonons in silica? One answer that immediately comes to mind is point defects. 

Indeed, point defects have been found to contribute significantly to the thermal conductivity of  some crystalline materials~\cite{Callaway1960}, a fact that is actively researched within the framework of thermoelectric  materials~\cite{Qin2021}. Glassy silica is well know to have many point defects~\cite{Pacchioni2012}, which are a significant concern for fiber optic applications~\cite{LoPiccolo2021}, where defect absorption spectrum appears to be the major source of information. A recent review by Jackson and Ballato~\cite{Jackson2025} emphasizes the relation between glass defect science and modern problems of high-power visible light generation and transportation in silicate glass optical fibre. A concern for point defects in silica has also arisen from microelectronics: El-Sayed et al.~\cite{ElSayed2015} have emphasized the role of defects in glassy SiO$_2$, particularly of their interaction with hydrogen, as may happen during the fabrication process of electronic devices, and their properties have been investigated using ab-initio numerical methods by Wilhelmer et al.~\cite{Wilhelmer2022}. An additional source of information is the role of point defects in the dielectric breakdown of amorphous films~\cite{Padovani2024}. From yet another direction,Yuan and Huang~\cite{Yuan2014} have highlighted the role of 5-fold Si coordination defects in the mechanical properties of glassy silica, particularly in its brittle to ductile transition, the importance of which has been put in perspective by Wondraczeck et al.~\cite{Wondraczek2022}.

Sushko et al.~\cite{Sushko2005} have sought to characterize defects in amorphous silica using classical molecular dynamics, and they highlight the difficulty of precisely defining what a ``defect'' is, in a random structure.  They identified, as a key unsolved issue the ``...accurate prediction of the relative concentration of different [point] defect types.'' To this day, there appears to be no reliable quantitative estimate of the number of point defects in glassy silica. Wimmer~\cite{Wimmer2017} quotes the value 10$^{18}$ --- 10$^{19}$ cm$^{-3}$, inferring it from measurements of  Kaczer~\cite{Kaczer2010} in two dimensional samples. This number ought to be a lower bound, given that is is tied to a specific physical behavior.

Point defects do not show up in the BP, as they possess no internal degrees of freedom to which a frequency can be associated. They can also account for the fact that we have neglected to give dynamics to the endpoints of the strings in our description. Therefore, a precise quantitative description of the thermal conductivity of silica will require to take these into account. Conservatively, we can think of the point defect density that we will introduce as a quantitative measure of the uncertainty in our modeling.

In what follows, we discuss the inclusion of point defects in our theoretical framework. With these results in hand, we will proceed to i) fix the value of $g$ in silica by fitting to speed of sound data, and then ii) determine the number of point defects needed in order to describe the thermal conductivity of silica.

\subsubsection{Point defects}

In what follows, we discuss how the properties of point defects, such as mass or number density, enter the phonon self-energies. We will do so from two perspectives: first, from Klemens' calculation~\cite{Klemens1955}, and second, from our own approach, detailed in Appendix~\ref{app:point-defect-scattering}. 

In either case, the net effect is the addition of phenomenological parameters $B_T$, $B_L$ to the phonon self-energies
\begin{align}
    \Sigma_T(\omega) &= \frac{\omega}{c_T} \frac{g^2}{4} \int_0^\infty \frac{dx}{x^2} \, G \! \left( \frac{x c_T}{\omega} \right) {\rm Tr}_{n_D} \!\! \left[ \frac{\tilde{F}_T(x) }{ D(x) - \tilde{T}(x) } \right] \nonumber  \\ & \quad + i B_T \omega^3  \, , \label{eq:SigmaT-B} \\
    \Sigma_L(\omega) &= \frac{\omega}{c_T} \frac{g^2}{4 \gamma^2} \int_0^\infty \frac{dx}{x^2} \, G \! \left( \frac{x c_T}{\omega} \right) {\rm Tr}_{n_D} \!\! \left[ \frac{\tilde{F}_L(x) }{ D(x) - \tilde{T}(x) } \right] \nonumber  \\ & \quad + i B_L \omega^3 \, , \label{eq:SigmaL-B}
\end{align}
accompanied by a frequency dependence of $\omega^3$. 
This reproduces the phonon relaxation time calculated by Klemens~\cite{Klemens1955}, corresponding to $\tau^{-1} \propto \omega^4$. The frequency dependence of such relaxation times is readily verified by setting $G = 0$ and using Eqs.~\eqref{eq:tauT-rel-optical-th} and~\eqref{eq:tauL-rel-optical-th}. Note that it does not affect the speed of sound.

The parameters $B_T$ and $B_L$ are not independent. However, depending on the nature of the microscopic defect, the relation between $B_T$ and $B_L$ may differ.
For example, if the impurity is a consequence of having an atom of different mass in its place, Klemens obtained
\begin{equation}
    B_L = B_T/\gamma^3 \, .
\end{equation}
with $B_T$ determined by
\begin{equation}
    B_T = \frac{1}{4\pi} \left( \frac{\Delta M}{M} \right)^2 \frac{n_P a^6 }{c_T^3}
\end{equation}
where $M$ is the mass of a constituent (non-defect) atom, $\Delta M$ is the mass difference of the defect, $a^3$ the unit cell volume, and we have introduced a point defect number density $n_p$ in place of the $1/G$ factor in the original formula\cite{Klemens1955}. Klemens derived a similar expression for a defect characterized by different elastic binding parameters. We note that while the expression for $B_T$ is written in terms of $a^3$, a variable that is natural for a crystalline solid, but not well-defined for a glassy material, a conversion to a natural variable in the continuum can be made by identifying $a^3 = M/\rho$, with $M$ the mass of the constituents of the glass.

These expressions were derived for scattering of phonons by a single impurity, and extrapolated to a density by multiplying the result such as to rescale it to account for the total number of defects. There is another possibility, however, that we describe in detail in Appendix~\ref{app:point-defect-scattering}, where a density of point defects is included in the continuum theory Lagrangian density. This leads to
\begin{equation}
    B_T = B_L \equiv B
\end{equation}
where
\begin{equation}
    B = \frac{2 + \gamma^{-3} }{12\pi} \frac{n_P M^2 }{\rho^2 c_T^3} \, , \label{eq:B-nP-rel}
\end{equation}
which can be interpreted as the average of Klemens' result over phonon polarizations, with a defect mass characterized by $\Delta M = M$. As the latter model fits naturally in our theoretical framework, we will use this one instead of using Klemens' model.

As it stands, we have three free parameters: $k_{\rm cut}$, $g$, and the newly introduced $B$. To describe the thermal conductivity and speed of sound, we will first proceed to fix $g$ so as to obtain a reasonably precise fit of the speed of sound in silica. Second, in the thermal conductivity calculation, we will fix $k_{\rm cut} = \pi/a$ by hand to values around $a \approx 0.5$ nm, and then extract $B$ by optimizing the rest of the fit. We will study the sensitivity of our results to varying $a$ and $B$.

\subsubsection{The speed of sound in Silica}

\begin{figure}
    \centering
    \includegraphics[width=0.95\linewidth]{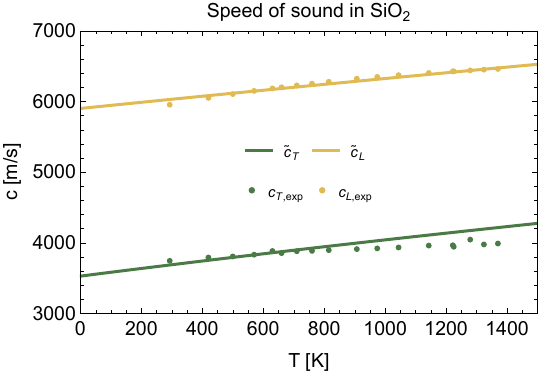}
    \caption{Fit of speed of sound data obtained by Polian et al.\cite{Polian2002} using Eqs.~\eqref{eq:cT-phys} and~\eqref{eq:cL-phys}. The resulting fit parameters are $c_L = 7275 \, {\rm m/s}$, $\gamma = 1.4263$ (corresponding to a bare speed of sound for shear waves of $c_T = 5100 \, {\rm m/s}$), and a phonon-dislon coupling of $g = 0.8112$.}
    \label{fig:speed-of-sound}
\end{figure}

In the same way as we did for Glycerol, we can use Eqs.~\eqref{eq:cT-phys} and~\eqref{eq:cL-phys} to fit $g$, $c_T$ and $c_L$ to match the renormalized speeds of sound $\tilde{c}_{T/L}$. This time we do not have the freedom to choose one of them by tuning the others, because the temperature dependence of the physical speed of sound provides many more data points than for Glycerol. The result of doing so is given in Figure~\ref{fig:speed-of-sound}, where we obtained
\begin{align}
    c_L^{{\rm SiO}_2} &= 7275 \, {\rm m/s} \, , \\
    \gamma_{{\rm SiO}_2} &= 1.4263 \, , \\
    g_{{\rm SiO}_2} &= 0.8112 \, .
\end{align}
We emphasize that $c_L$ is the ``bare'' longitudinal velocity of phonons, before accounting for interactions with line defects in the material. Only after doing so, the speed of propagation can be thought of as physical.

As we can see from Figure~\ref{fig:speed-of-sound}, the speed of sound of silica is well-described across a broad temperature range by selecting these three parameters in this way. Crucially, the temperature dependence of the speed of sound in this model is fully determined by the temperature dependence of the distribution of line defects (determined in turn by the BP density of states). In other words, we can view the determination of the bare quantities $c_L$, $\gamma$ (equivalently $c_T$) as fixing their values from speed of sound data at a single temperature, and the temperature dependence of the speed of sound as providing a direct diagnosis of the presence of line defects in the material.
Somewhat surprisingly, our results indicate that the bare Lam\'e parameter $\lambda$ is very close to vanishing. We will comment on the implications of this in the Discussion section.

\subsubsection{The thermal conductivity of Silica}

We can estimate the microscopic cutoff of glassy SiO${}_2$ by using the size of the unit cell in its crystalline form, which is roughly $a_{{\rm SiO}_2} = 0.5$ nm.
Correspondingly, we will set the momentum space cutoff to be $k_{\rm cut} = \pi/ a \approx 6.28 \, {\rm nm}^{-1}$, and also explore variations to $a = 0.45$ nm and $a = 0.55$ nm to test the sensitivity of our results to the value of the cutoff.

\begin{figure}
    \centering
    \includegraphics[width=0.95\linewidth]{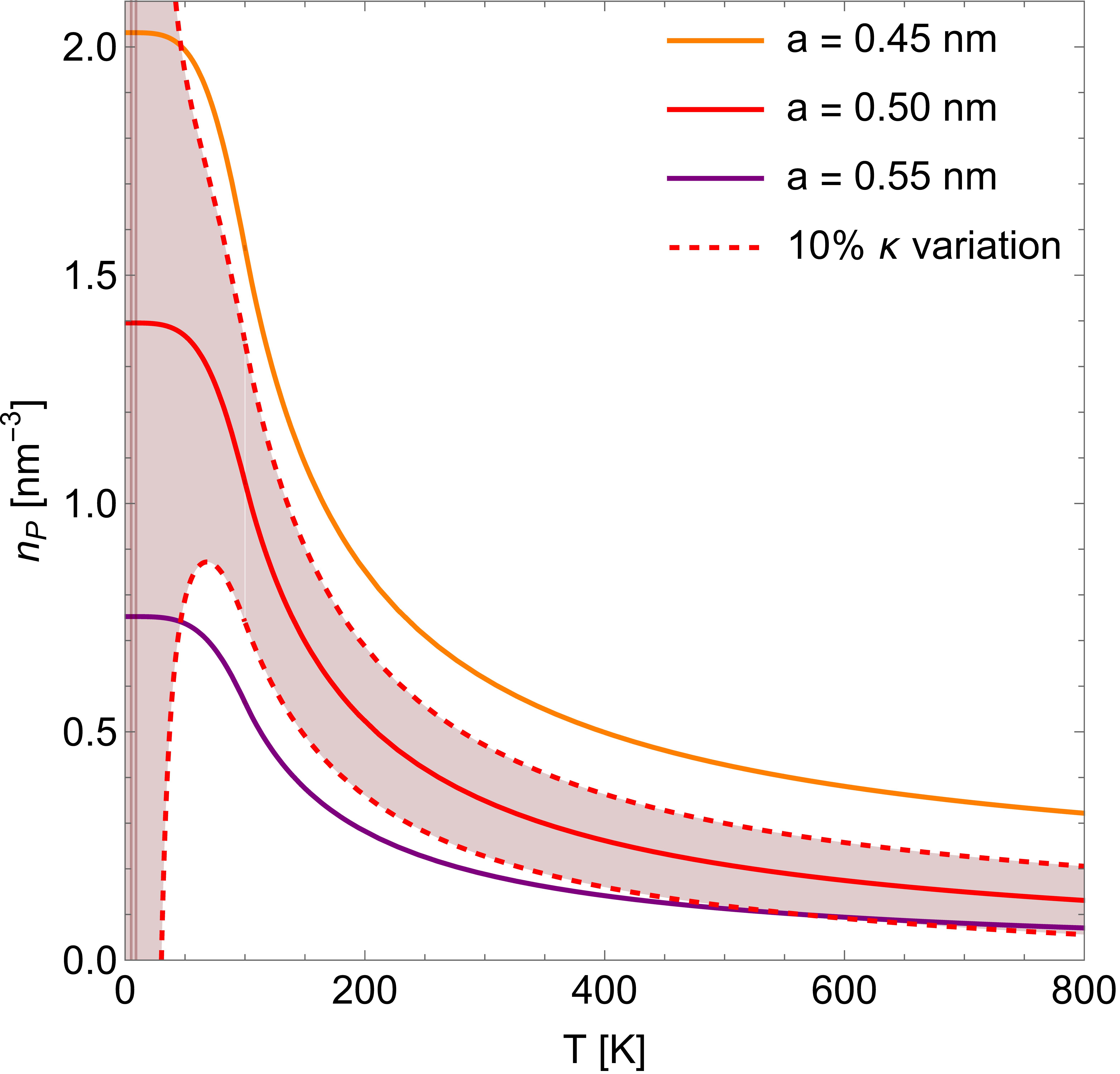}
    \caption{Result for the defect density $n_P(T)$ for SiO${}_2$ as described in the main text, for different values of the microscopic cutoff length $a_{{\rm SiO}_2}$. The fit parameters obtained were: i) $b_0 = 1.41 \times 10^{-43} \, {\rm s}^2$ and $b_1 = 1.38 \times 10^{-40} \, {\rm K} \, {\rm s}^2$ for $a_{{\rm SiO}_2} = 0.45$ nm, ii) $b_0 = 0 \, {\rm s}^2$ and $b_1 = 1.02 \times 10^{-40} \, {\rm K} \, {\rm s}^2$ for $a_{{\rm SiO}_2} = 0.50$ nm, and iii) $b_0 = 0 \, {\rm s}^2$ and $b_1 = 5.5 \times 10^{-41} \, {\rm K} \, {\rm s}^2$ for $a_{{\rm SiO}_2} = 0.55$ nm. Dashed lines: defect density needed to modify the thermal conductivity by $ 10\% $ for $a = 0.50$ nm.}
    \label{fig:BofT}
\end{figure}

\begin{figure}
    \centering
    \includegraphics[width=0.95\linewidth]{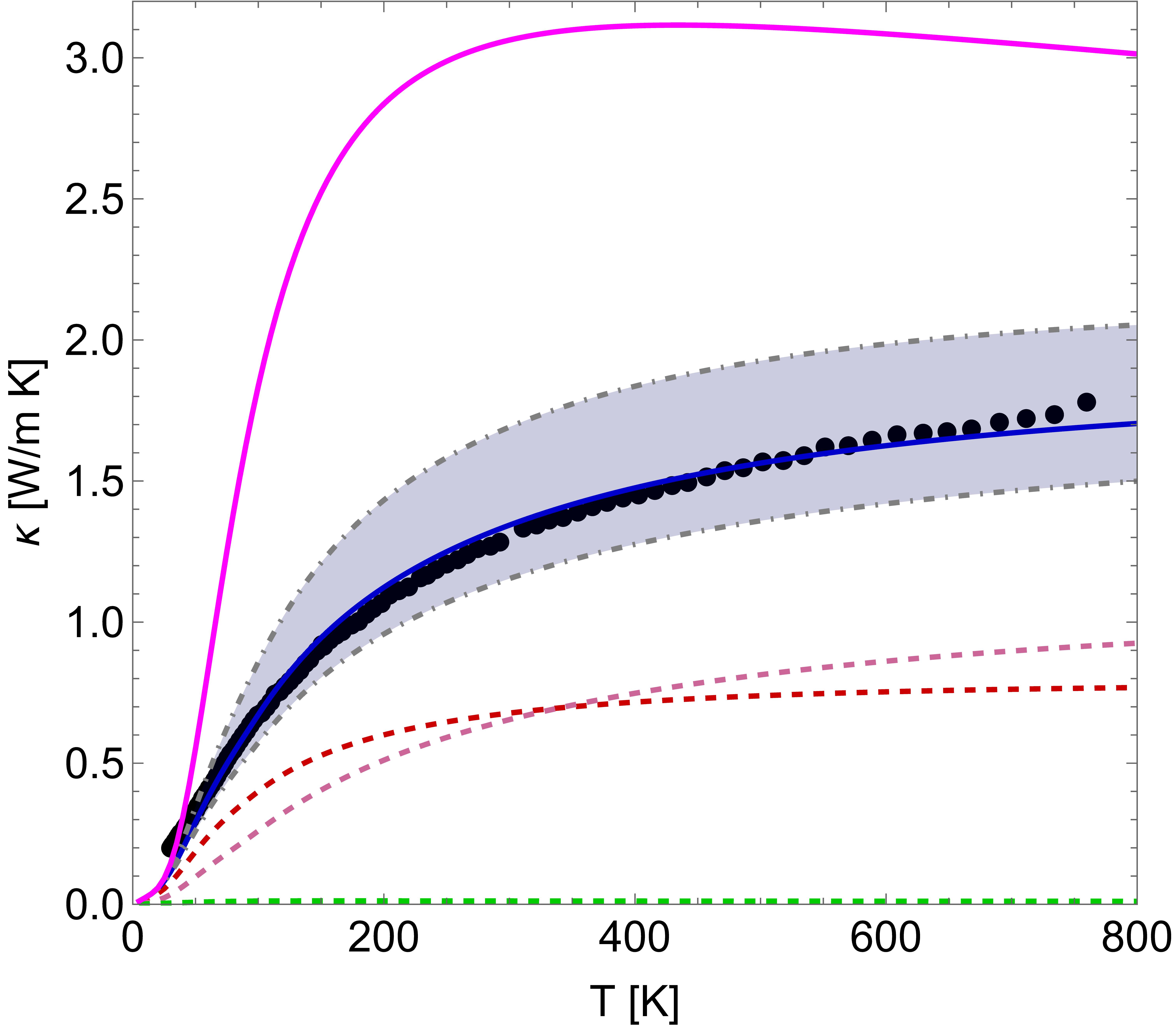}
    \includegraphics[width=0.95\linewidth]{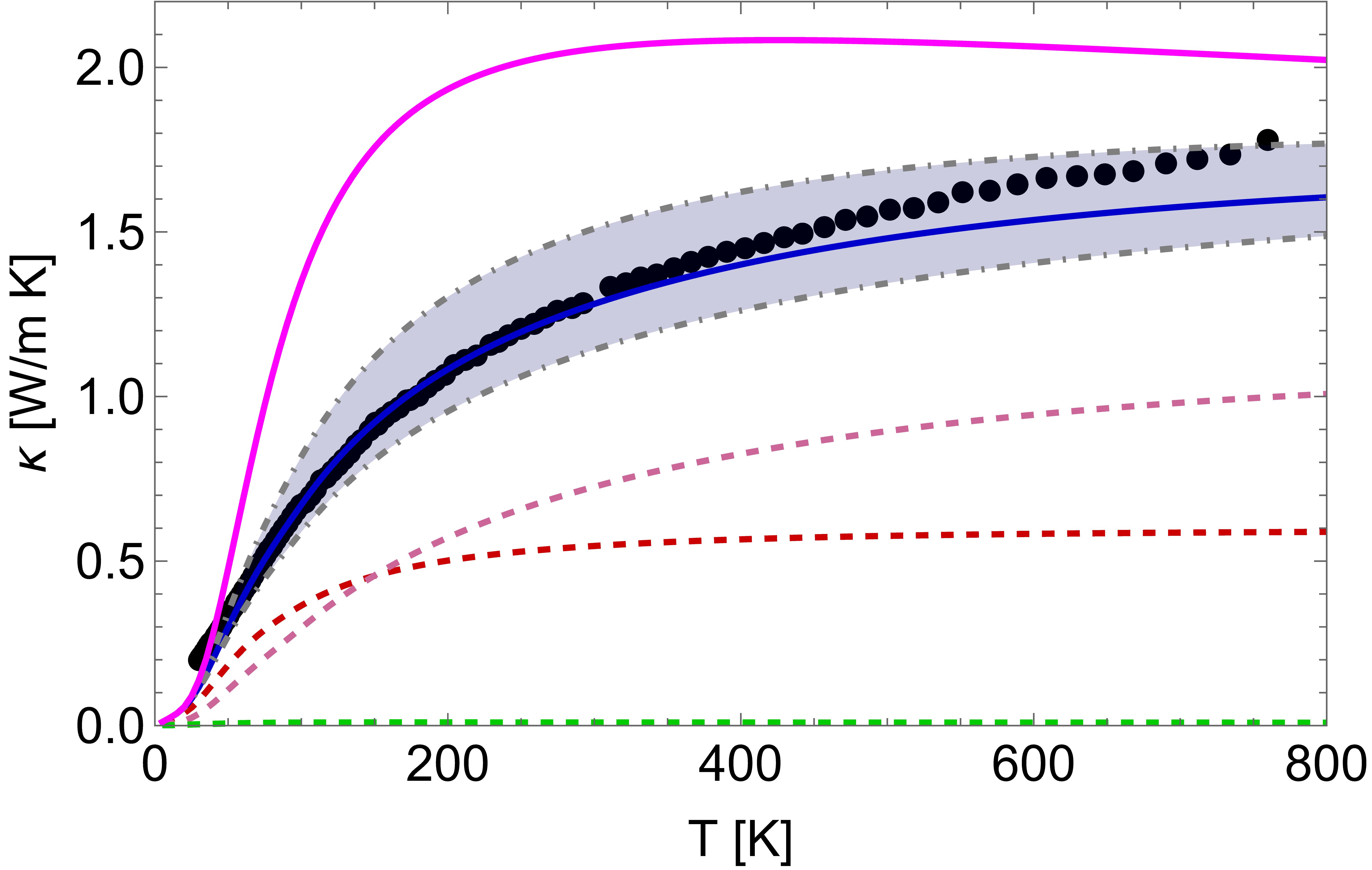}
    \includegraphics[width=0.95\linewidth]{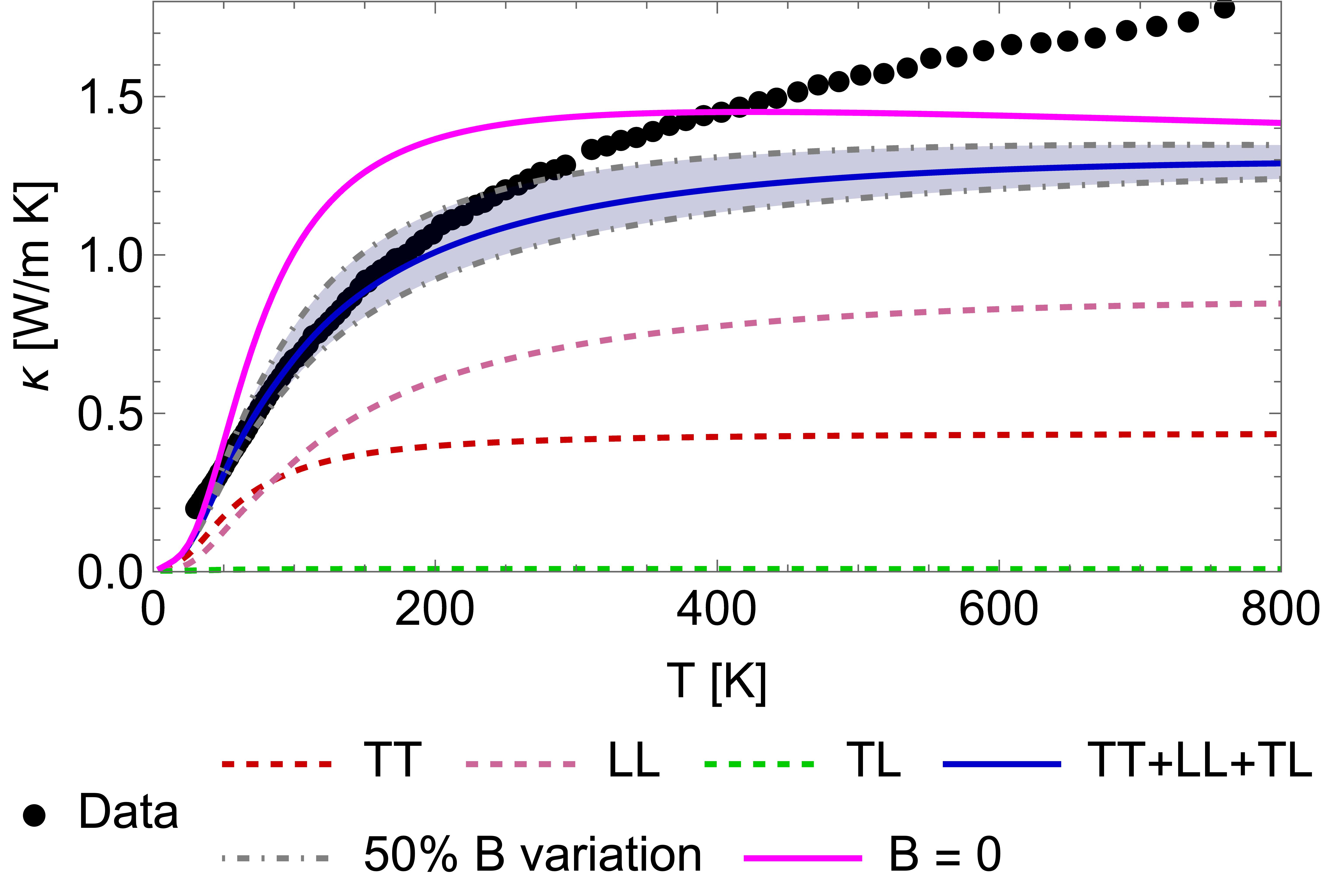}
    \caption{Comparison of the experimentally measured SiO${}_2$ thermal conductivity\cite{Cahill1990} (black circles) and the model developed in this work (blue curve), for different values of the microscopic cutoff length $a_{{\rm SiO}_2}$. In dashed curves, we show separately the contributions to the result from transverse modes (TT, red), longitudinal modes (LL, pink), and the interference between the two (TL, green), corresponding to each contribution in the second line of Eq.~\eqref{eq:conductivity}. In dot-dashed curves, we show how our model curves change by changing the value of the $B(T)$ parameter by 50\% (gray). Top panel: $a_{{\rm SiO}_2} = 0.45$ nm. Middle panel: $a_{{\rm SiO}_2} = 0.50$ nm. Bottom panel: $a_{{\rm SiO}_2} = 0.55$ nm. The magenta curve gives the theoretical  result phonon-strings interaction only, without point defects. The latter appear to be critical to account for the monotonous increase in the thermal conductivity of glassy silica in the temperature interval considered.}
    \label{fig:SiO2-cond}
\end{figure}

We can then fit $B(T)$ so as to fit thermal conductivity data. Since it is not within the scope of this work to study how $B$ may depend on temperature, we take a phenomenological model for $B(T)$ of the form
\begin{equation}
    B(T) = \begin{cases} b_0 + \frac{b_1}{T}  & T > 100 \, {\rm K} \\ b_0 + \frac{b_2 }{T^4 + b_3}   & T \leq 100 \, {\rm K} \end{cases} \, ,
\end{equation}
where we adjust the coefficients $b_1, b_0$ to describe the data, and fix $b_2,b_3$ imposing continuity of $B(T)$ and its first derivative at $T = 100$ K. In doing this, we use $M = 60$ a.u., the weight of a single SiO${}_2$ unit. The resulting $B(T)$ obtained in this way is given, implicitly, by means of Eq.~\eqref{eq:B-nP-rel} in Figure~\ref{fig:BofT}, where we display the point defect density needed to account for the thermal conductivity of SiO${}_2$. Finally, we present the thermal conductivity obtained with all of the above ingredients in Figure~\ref{fig:SiO2-cond}.

We see from Figure~\ref{fig:BofT} that, while cutoff-dependent, the resulting density of point defects is rather reasonable, being generally below 1 defect per cubic nanometer at 100 K and above. 
For lower temperatures, the contribution from line defects to the conductivity dominates to such an extent that the fit does not provide a practical constraint on the density of point defects, as illustrated by the shaded red band in the figure. The cutoff dependence of the value of $B$ is naturally understood when one recalls the fact that this is a continuum theory, in contrast to the microscopic (atomic) physics of the glass constituents. To match physical observables to the microscopic theory, a cutoff in the continuum theory is necessary. In fact, viewed from a Wilsonian perspective, so long as the physical processes of interest are below the scale of the cutoff, any value of the cutoff is valid provided the values of the couplings in the interaction terms in the continuum theory are suitably renormalized as a function of the cutoff. Thus, all parameters of the continuum theory are in principle tied to the value of the cutoff --- as $B$ is in order to match the physical thermal conductivity. On the other hand, because of how it affects the speed of sound, $g$ is fixed in a cutoff-independent way.\footnote{Having said this, it is worth to emphasize that we have \textit{not} explored the properties of our continuum theory under any kind of renormalization group flow, as would be necessary in order to have a precise understanding of the cutoff dependence of the parameters in this theory. We leave a study of these properties to future work.}

In light of this discussion, the interpretation of $B$ is a bit less straightforward than Klemens' and our derivations would suggest at a first glance: while it is reasonable to (roughly) interpret this parameter directly in terms of the defect density, the fact that parameters in a continuum theory intrinsically depend on its cutoff suggests that one should interpret
the value of $B(T)$ more loosely, accounting for all forms of disorder other than line defects in the continuum description of the material --- thus fully exploiting the effective (continuum description) nature of this parameter.

\section{Discussion and Conclusions}

In this paper we have adopted an approach to the thermal conductivity of glasses based on continuum mechanics: It ignores, from the start, the atomic structure of each specific glass. This approach has the advantage that it applies, by construction, to any and all glasses and thus can explain, again by construction, their commonalities. However this strength is, at the same time, its main weakness. The influence of a specific atomic structure has to be introduced by hand. %\sout{It is interesting that} 
However, as has been described in the text and will be discussed further in what follows, it is enough to include {\it one} material-dependent constant, a short distance cut-off $a$. The additional, glass-specific information that we introduce within a continuum framework is the presence of a boson peak, a quantity that can be independently determined by experiment. A basic input in our approach is to model this feature as a set of elastic, vibrating, strings that couple to the continuum phonons via the Peach-Koehler force (i.e., these strings are quantum Volterra dislocations).

The continuum theory of strings and phonons we employ has a second undetermined dimensionless parameter which we have called $g$: its indeterminacy is related to the fact that our strings are infinitely thin and a cut-off radius must be specified~\cite{Lund1988}. This parameter $g$, however, appears explicitly in the coupling between strings and phonons, a coupling that renormalizes the speed of propagation of shear and pressure waves. 

We have proposed that the basic physics of thermal transport in glasses above the plateau is that energy is carried by (continuum) phonons and that their transport is impeded by interactions with the elastic strings that are responsible for the presence of the boson peak. This same approach was used, within a classical field theory approach, to explain acoustic attenuation in glassy silica and glycerol~\cite{Bianchi2020}. Now we use a quantum field theory approach
where a Kubo formula determines the thermal conductivity. 

Depending on the richness of the data that is to be explained by the model, we have used three levels of complexity for the  implementation of our approach. The first is to explain the behavior of the thermal conductivity of four glasses in the temperature range 20-300 K, illustrated in Figure \ref{fig:conductivity-simple}.
At these low, but still above the plateau, temperatures, the  thermal conductivity initially rises linearly with temperature and then saturates for three of them. The fourth, silica, does not saturate but its rate of increase slows down. These behaviors can be quantitatively understood, in the framework of our model, as follows: at low temperatures, the thermal behavior will be dominated by low frequencies and long strings. From the data for silica and glycerol~\cite{Bianchi2020}, we extract the string length behavior for long lengths to be $p(L) \propto L^{-5}$. 
Taking a ballistic transport limit this proportionality constant 
is specified by the slope of the thermal conductivity curve around 20 K, and the resulting fits of Figure \ref{fig:conductivity-simple} with the values of Table \ref{tab:fit-params} are obtained.
Thus, the present formulation provides a quantitative answer to the following long-standing question (see e.g. Zeller and Pohl~\cite{Zeller1971}): Why is the thermal conductivity of a glass an increasing function of temperature in a range of temperatures where the thermal conductivity of the corresponding crystal is a decreasing function? In the language of Allen et al.~\cite{AllenFeldman1999}, said thermal conductivity is provided by ``propagons'' (our continuum mechanics renormalized phonons) that are scattered by ``locons'' (our continuum mechanics elastic strings). In our case, the localized modes associated with the normal modes of a glass---they give rise to the boson peak---decay with distance as a power law~\cite{Bianchi2020}, in agreement with recent numerical simulations~\cite{Lerner2016,Gartner2016,Kapeteijns2018,Lerner2018,Shimada2018}, and not exponentially, as would be the case if they were Anderson-localized modes.

The second step in complexity is to use the full Eq. \eqref{eq:conductivity} for the thermal conductivity, for which the full string length distribution $p(L)$ is needed. This is what we have done in Section \ref{gly} for glycerol, with the result shown in Figure \ref{fig:Glycerol-cond}. A very good fit to the data is obtained with a short distance cut-off $a_{\rm Glycerol}\sim 0.7$ nm, a quite  reasonable value, and a string-phonon coupling constant $g_{\rm Glycerol} \sim 1$, determined by the experimentally measured (i.e., dressed) velocities of wave propagation. We have not found thermal conductivity data for glycerol below 50 K. It would be nice to measure it, and see whether it conforms to our theory.

Finally, in the case of silica, once the full density of states $p(L)$ is included in Eq. \eqref{eq:conductivity} and not just its long string approximation, the overall picture of (continuum) phonons scattered by dislons (quantum string oscillations)  provides a good fit to the experimental data for thermal conductivity in the range $\sim 30-100$ K with a suitable choice of the short-distance cutoff $a$ (not shown) but not at higher temperatures. 
What is to be done?
In this regime the thermal transport of crystals is dominated by the umklapp scattering of phonons~\cite{Ashcroft1976}. In the case of glasses there is no crystal structure hence no umklapp process can be defined. Point defects, however, have been found to contribute significantly in some crystalline materials~\cite{Callaway1960}, a fact that is actively researched within the framework of thermoelectric  materials~\cite{Qin2021}. Glassy silica is well know to have many point defects~\cite{Pacchioni2012}, which are a significant concern for fiber optic applications~\cite{LoPiccolo2021}.

It stands to reason that said point defects must interact with phonons and they will be an additional source of thermal resistance. In order to quantify this physics, we have added an appropriate term to our basic Lagrangian, as explained in detail in Appendix \ref{app:point-defect-scattering}. The price to pay is the introduction of a second, temperature dependent, parameter $B$ with units the cube of time, proportional to the product of the defect density $n_P$ and defect mass $M$. Taking $M$ as the  mass of a single SiO$_2$ molecule we find that, given the experimentally measured wave speeds and density of states for the Boson peak at various temperatures, the density of defects shown in Figure  \ref{fig:BofT} provides a very good fit to the thermal conductivity of glassy silica in the 30-750 K temperature range, with a very reasonable short-distance cut-off $a \sim 0.5$ nm.

The overall conclusion we draw from the work we have presented here is fairly simple: The basics of thermal transport in amorphous materials above the plateau can be understood if they are modeled as continuous media with randomly placed, and oriented, elastic strings that interact with phonons via the Peach-Koehler force. A detailed quantitative comparison with experimental data necessitates the implementation of a quantum field theory approach.  
The foregoing suggests several possible avenues for further research including, of course, the comparison with data for additional glasses, as well as the role of strings in other glass-related phenomena, and their precise relation to the material microscopic structure.

\begin{acknowledgments}
The work of BSH was supported in part by grant NSF PHY-2309135 to the Kavli Institute for Theoretical Physics (KITP), and by grant 994312 from the Simons Foundation. The work of FL was supported by Fondecyt grant 1230938. We thank A. Tanguy and N. Shchlebanov for useful discussions.
\end{acknowledgments}

\appendix

\section{Fit to the boson peak density of states} \label{app:DOS-fits} 

As described in the main text, upon neglecting the contribution from excited string states, the boson peak (BP) density of states $g_S(\omega_0)$ determines the length distribution of defects via
\begin{equation}
    p(L) = g_S(\omega_0(L)) \left|\frac{d\omega_0}{dL} \right| \, ,
\end{equation}
or, more explicitly,
\begin{equation}
    p(L) = \frac{\pi \alpha c_T}{L^2}  g_S \! \left( \frac{\pi \alpha c_T}{L} \right)  \, .
\end{equation}

That is to say, given the functional form of $g_S$, the length distribution is fully determined. However, given that the BP data measurements are discrete in frequency, many fitting models are possible for $g_S$. In fact, because $p(L)$ enters our results through integral expressions, the precise fitting form is not essential. Thus, we simply choose \"ansatze for $g_S$ that make the fitting procedure as simple as possible.

\begin{figure}
    \centering
    \includegraphics[width=0.95\linewidth]{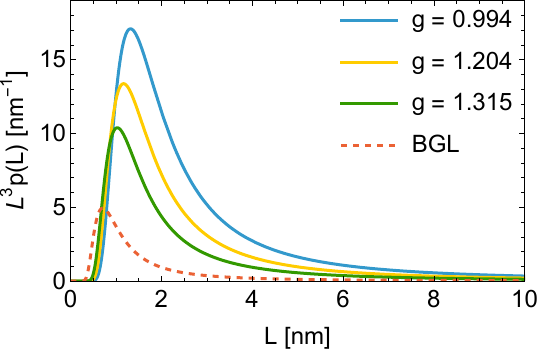}
    \caption{String length distribution in Glycerol determined from the BP density of states determined by BGL\cite{Bianchi2020} (see their Figure 1). For comparison, we display the length distribution BGL extracted from the same BP data, accounting for the factor of $2$ in the mode counting but maintaining the (mild) discrepancy between the dispersion relation of the string modes. We show the resulting length distribution for the different choices of the coupling $g_{\rm Glycerol}$ we explored in the main text.}
    \label{fig:pL-glycerol-new}
\end{figure}

For Glycerol, we use the same fitting form as BGL, i.e.,
\begin{equation}
    \frac{g_S(\omega)}{\omega^2} = q_0 \omega \exp \left( - q_1 (\omega - q_2)^2 \right) \, .
\end{equation}
We show the results of this fit in terms of $p(L)$ in Fig.~\ref{fig:pL-glycerol-new}.

The main reason for the large difference in $p(L)$ between our study and BGL is that the bare speeds of sound are significantly different from the physical ones in our approach. After going through the calculation, one can check that this difference implies that our determination of $p(L)$ is a factor of
\begin{equation}
    \left( \frac{\alpha}{\tilde{\alpha}} \frac{c_T}{\tilde{c_T} } \right)^{4} \, ,
\end{equation}
larger in amplitude, where 
\begin{equation}
    \tilde{\alpha} = \sqrt{2(1 - \tilde{c}_T^2/\tilde{c}_L^2 )/(1 + \tilde{c}_T^4/\tilde{c}_L^4)} \, .
\end{equation}

Our parametrization of the BP density of states $g_S(\omega)$ for SiO${}_2$ is of the form
\begin{equation}
    \frac{g_S(\omega)}{\omega^2} = \frac{a_1 \omega + a_3 \omega^3}{Q_5(\omega)}
\end{equation}
where $Q_5(\omega)$ is a polynomial of degree $5$ in $\omega$. While our choice is clearly somewhat arbitrary, its asymptotics at small and large $\omega$ accurately describe the data. In comparison with the BGL parametrization, they share the same low-frequency asymptotics, but differ at large $\omega$. While this has quantitative consequences for the thermal conductivity at high temperatures, the low-temperature regime, which is dominated by the low frequency behavior of $g_S(\omega)$, remains the same.

\section{Auxiliary expressions for the self-energy in the string model} \label{app:aux-expr}

The first aim of this Appendix is to show the explicit derivation going from Eqs.~\eqref{eq:selfenergyT-euclidean} and~\eqref{eq:selfenergyL-euclidean} to Eqs.~\eqref{eq:SigmaT} and~\eqref{eq:SigmaL}. The first step is to carry out the analytic continuation. Following~\cite{Lund2019}, we obtain
\begin{align}
    \Sigma_T(k,\omega) &= \frac{g^2 c_T^2}{4} \int dL \, L \, p(L) {\rm Tr}_{n_D} \!\! \left[ \frac{F_T(k,\omega)}{1 - T(\omega)} \right] \, , \label{eq:selfenergyT-app} \\
    \Sigma_L(k,\omega) &= \frac{g^2 c_T^2}{4 \gamma^2} \int dL \, L \, p(L)  {\rm Tr}_{n_D} \!\! \left[ \frac{F_L(k,\omega)}{1 - T(\omega)} \right] \, , \label{eq:selfenergyL-app}
\end{align}
where we have used that $\tfrac{\mu^2 b^2}{m \rho c_T^2} = g^2 c_T^2$.

In the approximation where we impose $k = k(\omega) = \omega/c_T$ in Eq.~\eqref{eq:selfenergyT-app} and $k = k(\omega) = \omega/c_L$ in Eq.~\eqref{eq:selfenergyL-app}, we can further write
\begin{align}
    &\Sigma_T(\omega) \label{eq:selfenergyT-app-1} \\
    &= \frac{g^2 c_T^2}{4} \int \! dL \, L \, p(L) \frac{L^2}{c_T^2} {\rm Tr}_{n_D} \!\! \left[ \frac{\tilde{F}_T(\omega L/c_T )}{D(\omega L/c_T) - \tilde{T}(\omega L/c_T)} \right] \, ,   \nonumber \\
    &\Sigma_L(\omega) \label{eq:selfenergyL-app-1}  \\
    &= \frac{g^2 c_T^2}{4 \gamma^2} \int \! dL \, L \, p(L) \frac{L^2}{c_T^2} {\rm Tr}_{n_D} \!\! \left[ \frac{\tilde{F}_L(\omega L/c_T)}{D(\omega L/c_T) - \tilde{T}(\omega L/c_T)} \right] \, ,  \nonumber 
\end{align}
with $\tilde{F}_{T/L},D,\tilde{T}$ defined below. The factors of $\tfrac{L^2}{c_T^2}$ appear when converting from the expressions in terms of $F_{T/L},T$ given in~\cite{Lund2019} to $\tilde{F}_{T/L},D,\tilde{T}$. Introducing the frequency-rescaled length variable $x = \omega L/c_T$, these expressions become
\begin{align}
    &\Sigma_T(\omega) \label{eq:selfenergyT-app-2} \\
    &=  \frac{g^2 \omega}{4 c_T } \int \frac{dx}{x^2} \left( \frac{x c_T}{\omega} \right)^5 p\!\left(\frac{x c_T}{\omega}\right) {\rm Tr}_{n_D} \!\! \left[ \frac{\tilde{F}_T(x )}{D(x) - \tilde{T}(x)} \right] \, , \nonumber \\
    &\Sigma_L(\omega) \label{eq:selfenergyL-app-2} \\
    &= \frac{g^2  \omega}{4 c_T \gamma^2} \int \frac{dx}{x^2} \left( \frac{x c_T}{\omega} \right)^5 p\!\left(\frac{x c_T}{\omega}\right)  {\rm Tr}_{n_D} \!\! \left[ \frac{\tilde{F}_L(x)}{D(x) - \tilde{T}(x)} \right] \, ,  \nonumber 
\end{align}
which are exactly Eqs.~\eqref{eq:SigmaT} and~\eqref{eq:SigmaL} after identifying $G(L) = L^5 p(L)$.

Equations~\eqref{eq:SigmaT} and~\eqref{eq:SigmaL} (later on in the text also Eqs.~\eqref{eq:SigmaT-B} and~\eqref{eq:SigmaL-B}, as well as Eqs.~\eqref{eq:selfenergyT-app-2} and~\eqref{eq:selfenergyL-app-2} in this Appendix) require the specification of functions not given explicitly in the main text. The second purpose of this Appendix is to do so. Explicitly, they are given by
\begin{widetext}

\begin{align}
    [D(x)]_{n n'} &= (n^2 \pi^2 \alpha^2 - x^2) \delta_{nn'} \, , \\
    [\tilde{F}_T(x)]_{n n'} &= n n' \pi^2 \int_{-1}^1 (1-u^4) \frac{1 - (-1)^{n} e^{i x u}  }{ (n \pi)^2 - (x u)^2 } \frac{1 - (-1)^{n'} e^{-i x u}  }{ (n' \pi)^2 - (x u)^2 } \, , \\
    [\tilde{F}_L(x)]_{n n'} &= 2 n n' \pi^2 \int_{-1}^1 (1-u^2)^2 \frac{1 - (-1)^{n} e^{i x u/\gamma}  }{ (n \pi)^2 - (x u/\gamma)^2 } \frac{1 - (-1)^{n'} e^{-i x u/\gamma}  }{ (n' \pi)^2 - (x u/\gamma)^2 } \, , \\
    [\tilde{T}(x)]_{n n'} &= \frac{i x^3}{\ell} \int_0^1 \!\! ds \! \int_0^s \!\! ds' \! \int_0^1 \!\! du (1 - u^2) \left[ \sin (n \pi s) \sin(n' \pi s') + (n \leftrightarrow n') \right] \nonumber \\
    & \quad \quad \quad \quad \quad \quad \quad \quad \quad \quad \quad \quad \quad \! \times \left[ (1 + u^2) e^{i(s-s')xu} + \frac{1-u^2}{\gamma^5} e^{i(s-s')x u/ \gamma} \right] \, .
\end{align}

\end{widetext}

Note that the value of $\gamma$ that enters these formulas is obtained from the ``bare'' Lagrangian parameters $\gamma = c_L/c_T$, not from the physical speed of sound of each material.

\section{Calculation of the thermal conductivity in the cases of Silica and Glycerol}

Given phonon self-energies $\Sigma_T$, $\Sigma_L$ as defined in the main text, e.g., in Eqs.~\eqref{eq:rhoT-expr} and~\eqref{eq:rhoL-expr}, which we assume depend only on frequency $\omega$ (in practice, filling in its $k$ dependence by making use of the corresponding bare dispersion relation $k = \omega / c$), it is possible to reduce the complexity of the integral expressions involved in calculating the thermal conductivity down to a single one-dimensional integral.

If one further introduces a rescaled frequency variable $y = \omega/T$, and defines
\begin{align}
    \sigma_T^2 &= 1 + \frac{\big( {\rm Im}\{\Sigma_T\} \big)^2}{\big(1 - {\rm Re}\{\Sigma_T\} \big)^2 } \, , \\
    \sigma_L^2 &= 1 + \frac{\big( {\rm Im}\{\Sigma_L\} \big)^2}{\big(1 - {\rm Re}\{\Sigma_L\} \big)^2 }  \, , \\
    \chi_T &= \frac{y T}{c_T \sigma_T} \frac{1}{\sqrt{1 - {\rm Re}\{\Sigma_T\}}} \, , \\
    \chi_L &= \frac{y T}{c_L \sigma_L} \frac{1}{\sqrt{1 - {\rm Re}\{\Sigma_L\}}} \, ,
\end{align}
then the thermal conductivity (in units where $\hbar = k_B = 1$) is given by
\begin{equation}
    \kappa = \frac{T^2}{3\pi^3} \left[ \frac{2\tilde{\kappa}_{TT}}{c_T} + \frac{\tilde{\kappa}_{LL}}{c_L} + \frac{(c_L^2 - c_T^2)^2}{(c_T c_L)^{5/2}} \tilde{\kappa}_{TL}  \right] \, ,
\end{equation}
where
\begin{widetext}
\begin{align}
    \tilde{\kappa}_{TT} &= \int_{-\infty}^{\infty} dy \frac{y^3 e^y}{(e^y - 1)^2 } \frac{\sigma_T^2 - 1}{\big(1 - {\rm Re}\{\Sigma_T\} \big)^{5/2} \sigma_T^5  } \int_0^{k_{\rm cut}/\chi_T} \frac{u^8 du}{[u^4 - 2u^2 + \sigma_T^2]^2} \, , \\
    \tilde{\kappa}_{LL} &= \int_{-\infty}^{\infty} dy \frac{y^3 e^y}{(e^y - 1)^2 } \frac{\sigma_L^2 - 1}{ \big(1 - {\rm Re}\{\Sigma_L\} \big)^{5/2} \sigma_L^5 } \int_0^{k_{\rm cut}/\chi_L} \frac{u^8 du}{[u^4 - 2u^2 + \sigma_L^2]^2} \, , \\
    \tilde{\kappa}_{TL} &= \int_{-\infty}^{\infty} dy \frac{y^3 e^y}{(e^y - 1)^2 } \sqrt{\frac{\sigma_T^2 - 1}{ \big(1 - {\rm Re}\{\Sigma_T\} \big)^{5/2} \sigma_T^5} \frac{\sigma_L^2 - 1}{ \big(1 - {\rm Re}\{\Sigma_L\} \big)^{5/2} \sigma_L^5} } \nonumber \\ & \quad \quad \quad \quad \,\, \times \int_0^{\frac{k_{\rm cut}}{\sqrt{\chi_T \chi_L}} } \frac{u^8 du}{\left[ \frac{\chi_L^2}{\chi_T^2} u^4 - 2 \frac{\chi_L}{\chi_T} u^2 + \sigma_T^2 \right] \left[ \frac{\chi_T^2}{\chi_L^2} u^4 - 2 \frac{\chi_T}{\chi_L} u^2 + \sigma_L^2 \right]} \, .
\end{align}
\end{widetext}
The integrals over $u$ may be done analytically, although this is a task better suited for symbolic integration software.

\section{Scattering of long-wavelength phonons by many point defects} \label{app:point-defect-scattering}

In order to supplement our model, for the case of silica we also consider the presence of point defects. In this appendix, we give a treatment of the effects of point defects on phonon propagation on the same footing as for linear defects.

The starting point is to consider an isolated point defect, which, we shall assume, describes a particle of mass $M$ that does not form part of the rest of the continuum, whose Lagrangian is given by
\begin{align} \label{eq:point-defect-lagrangian}
    L_{\rm defect} &= \frac{1}{2} M \dot{Y}_i \dot{Y}_i - \frac{1}{2} M f^2 Y_i Y_i \\ &\quad + \frac{M}{\rho} \left( \mu \frac{\partial^2 u_i({\bf x}_0)}{\partial x_k \partial x_k} Y_i + (\lambda + \mu) \frac{\partial^2 u_k({\bf x}_0)}{\partial x_k \partial x_i} Y_i \right) \nonumber
\end{align}
where $Y_i$ is the displacement of the defect (located at position ${\bf x}_0$), $M$ is the mass of the defect, $f$ is the frequency associated to a restoring force that keeps the defect around its equilibrium position (we will set it to zero later on; we will keep it for the moment for the sake of generality), and we assume that its interaction with the {elastic continuum} is given by the force any {volume} cell would experience due to the strain
tensor of the {surrounding continuum}, given explicitly by the second line in the equation above.

The two-point function of the defect in imaginary time (i.e., its correlation function with itself at some temporal separation) can be written in terms of Matsubara frequency modes as
\begin{align}
    \Delta_{Y}^{E}(k_n) &\equiv \int_0^\beta d\tau e^{i k_n \tau} \langle Y_i(\tau) Y_j(0) \rangle \nonumber \\  &= \frac{\delta_{ij} \hbar/M}{k_n^2 + f^2 - \Pi(k_n)} \, .
\end{align}

As it turns out, $\Pi$ can be calculated directly using the free phonon propagator as a mediator that allows the defect to absorb/emit energy. It is given by
\begin{equation}
    \Pi = \frac{M}{3\rho c_T^3} \left( 2 + \gamma^{-3} \right) \int \frac{d\nu}{ 2\pi^2 } \frac{\nu^6}{k_n^2 + \nu^2} \, .
\end{equation}
The last integral is badly divergent in the continuum. Upon the introduction of a UV cutoff $\Lambda$, it is of the form
\begin{equation}
    \int \frac{d\nu}{ 2\pi^2 } \frac{\nu^6}{k_n^2 + \nu^2} = a_5 \Lambda^5 + a_3 \Lambda^3 k_n^2 + a_1 \Lambda k_n^4 + F(k_n) \, ,
\end{equation}
where $F(k_n)$ is finite in the continuum limit. It is not hard to verify that (e.g., by taking $3$ derivatives with respect to $k_n^2$ and then calculating the integral)
\begin{equation}
    F(k_n) = - \frac1{4\pi} \left(k_n^2\right)^{5/2} \, .
\end{equation}
Upon analytically continuing $\Delta_Y^E$ to obtain a retarded propagator, the only piece that will contribute to dissipation is the imaginary part, which can only be generated by odd powers of $k_n$. This is why the calculation we just carried out (even if it is, strictly speaking, divergent) contains physically meaningful information.

The constants $a_5 \Lambda^5$ and $a_3 \Lambda^3$ can formally be absorbed into redefinitions of the ``bare'' parameters $M$ and $f$, thus renormalizing them. The term proportional to $a_1$ is actually problematic in that it generates a term that was not present to begin with. This means one should in general write
\begin{align}
    \Delta_{Y}^{E}(k_n) = \frac{\delta_{ij} \hbar/M}{r_4 k_n^4 + r_2 k_n^2 + r_0 f^2 + \frac{2 + \gamma^{-3} }{12 \pi} \frac{M}{\rho c_T^3} (k_n^2)^{5/2} } \, ,
\end{align}
and fit the parameters $r_4$, $r_2$, $r_0$ by matching to data.

However, our approach towards dealing with these parameters will be rather practical: in the spirit of having a defect propagator that describes the physics of a massive particle coupled to an elastic continuum which is otherwise free, we will simply set $r_4 = r_0 = 0$ and $r_2 = 1$. As such, we will use
\begin{align}
    \Delta_{Y}^{E}(k_n) = \frac{\delta_{ij} \hbar/M}{ k_n^2 + \frac{2 + \gamma^{-3} }{12 \pi} \frac{M}{\rho c_T^3} (k_n^2)^{5/2} } \, ,
\end{align}
as our model for the dynamics of an individual point defect in an elastic continuum.

The next step in order to make contact with the macroscopic description of a material is to introduce a continuous distribution of these point defects. That is to say, adding a label ${\bf x}_0$ to each $Y_i$ displacement, and writing
\begin{equation}
    L = L_{{\rm ph}+{\rm dislocs}} + n_P \int d^3{\bf x}_0  L_{{\rm defect}, \, {\bf x}_0} \, .
\end{equation}
A straightforward calculation then leads to a phonon self-energy with two contributions
\begin{equation}
    \Sigma_{T/L} = \Sigma_{T/L}^{\rm dislocs} + \Sigma_{T/L}^{\rm points} \, ,
\end{equation}
the first of which ($\Sigma_{T/L}^{\rm dislocs}$) we discussed at length in the main text, and the latter is given by
\begin{align}
    \Sigma_{T}^{\rm points} &=  \frac{n_P M}{\rho} \frac{ c_T^2 k^2 }{k_n^2 + \frac{2 + \gamma^{-3} }{12 \pi} \frac{M}{\rho c_T^3} (k_n^2)^{5/2} } \, , \label{eq:sigmaTpointsApp} \\
    \Sigma_{L}^{\rm points} &= \frac{n_P M}{\rho} \frac{ c_L^2 k^2 }{k_n^2 + \frac{2 + \gamma^{-3} }{12 \pi} \frac{M}{\rho c_T^3} (k_n^2)^{5/2} } \, . \label{eq:sigmaLpointsApp}
\end{align}

\begin{figure*}
    \centering
    \includegraphics[width=0.49\linewidth]{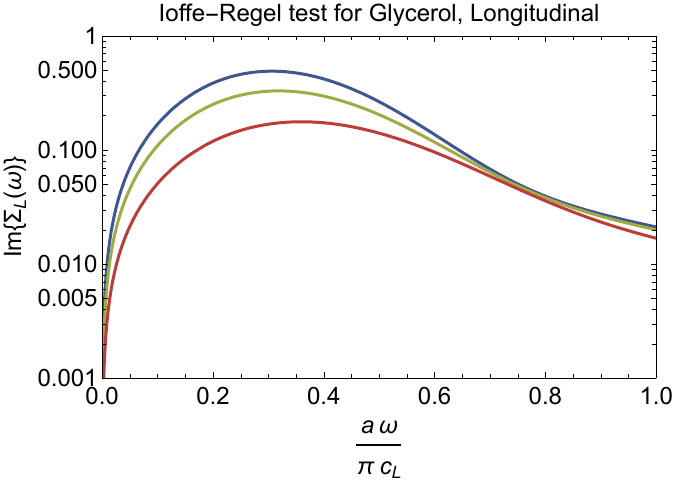}
    \includegraphics[width=0.49\linewidth]{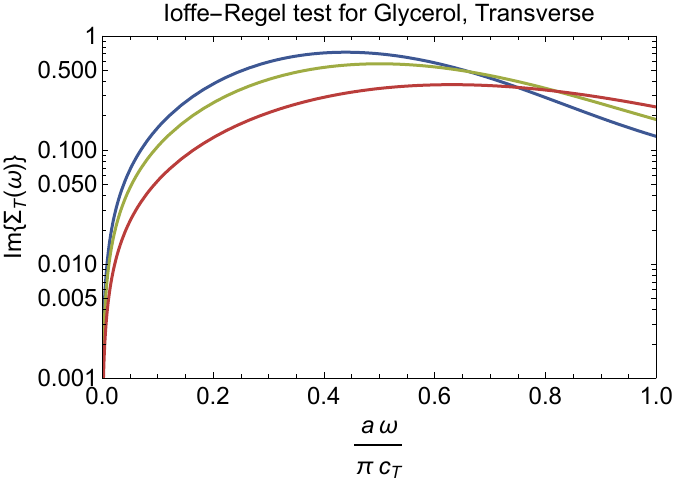}
    \includegraphics[width=0.37\linewidth]{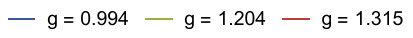}
    \caption{Imaginary part of the dimensionless longitudinal self-energy $\Sigma_L$ (left column) and transverse self-energy $\Sigma_T$ (right column) in Glycerol for the three values of the coupling constant $g$ considered in the main text. Top row: $g_{\rm Glycerol} = 0.994$. Middle row: $g_{\rm Glycerol} = 1.204$. Bottom row: $g_{\rm Glycerol} = 1.315$. }
    \label{fig:Ioffe-Regel-Glycerol}
\end{figure*}

\begin{figure*}
    \centering
    \includegraphics[width=0.49\linewidth]{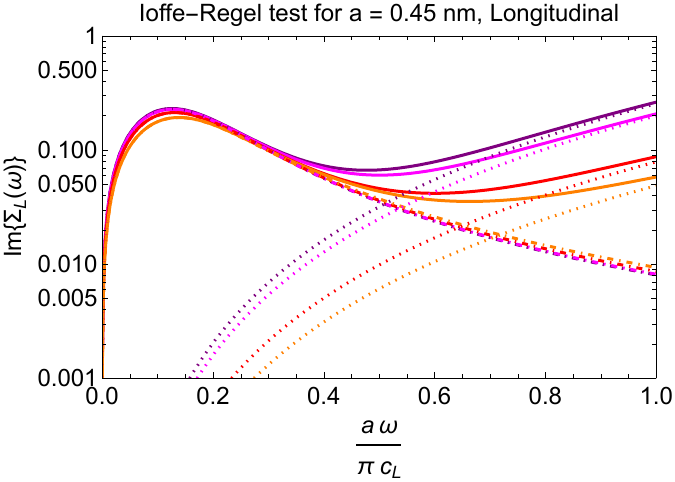}
    \includegraphics[width=0.49\linewidth]{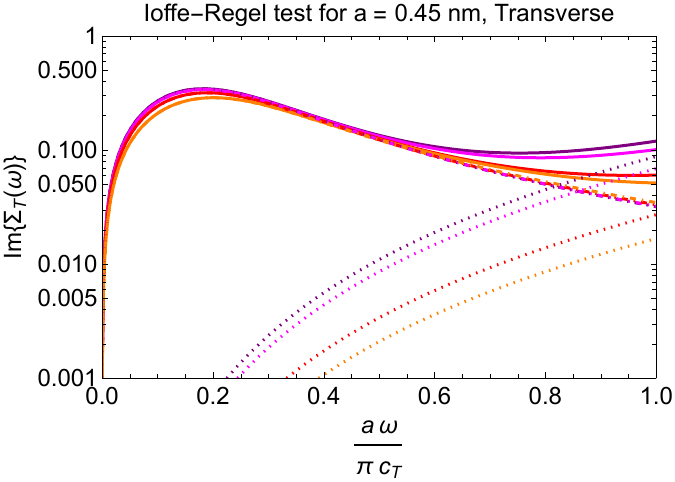}
    \includegraphics[width=0.49\linewidth]{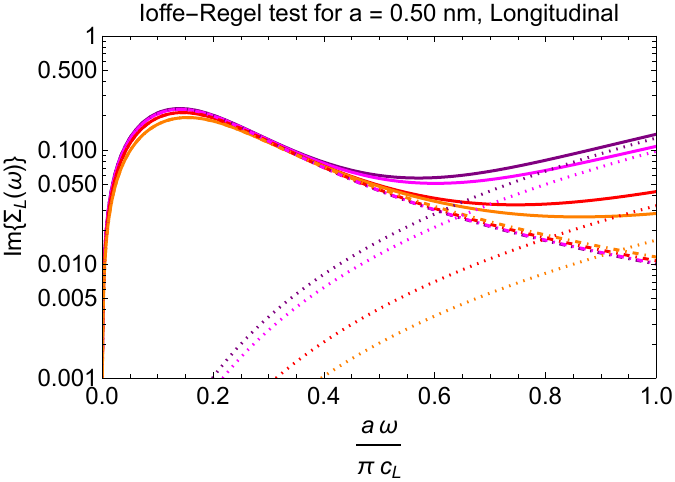}
    \includegraphics[width=0.49\linewidth]{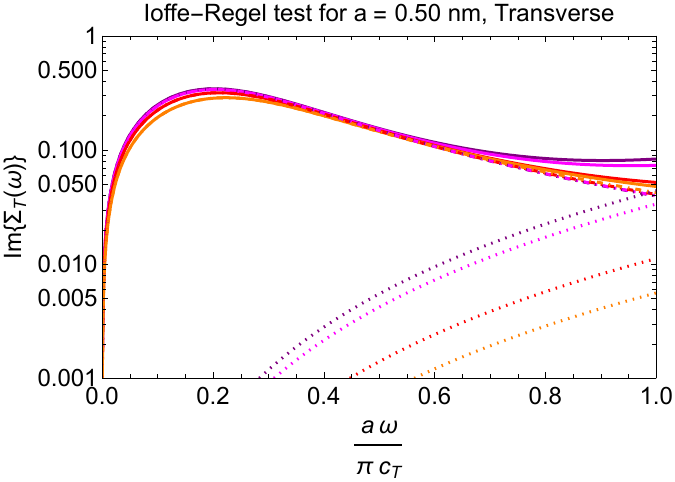}
    \includegraphics[width=0.49\linewidth]{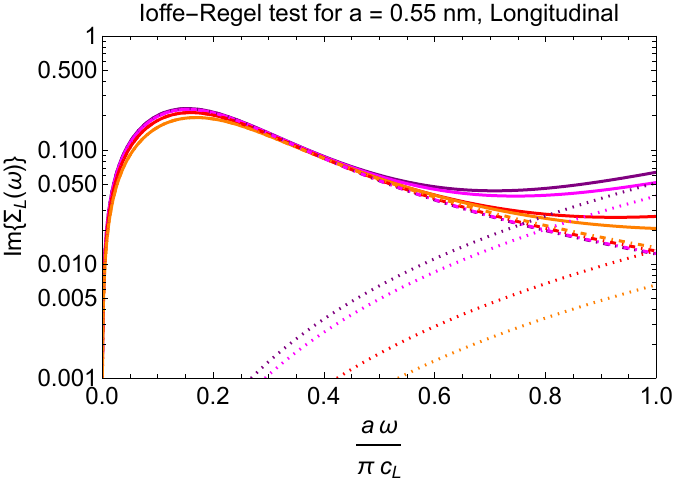}
    \includegraphics[width=0.49\linewidth]{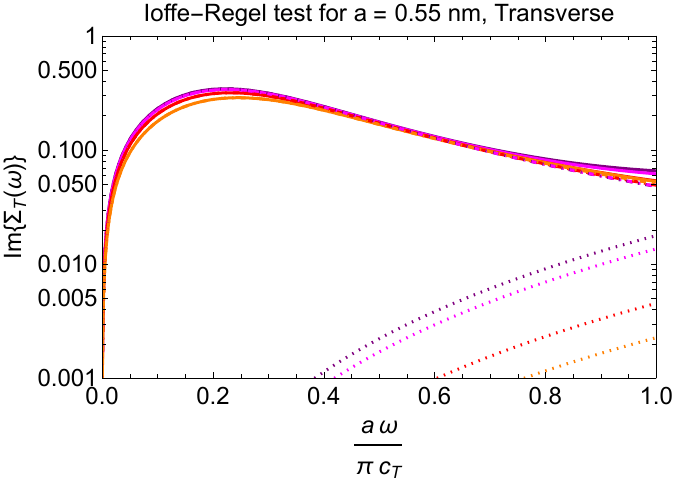}
    \includegraphics[width=0.95\linewidth]{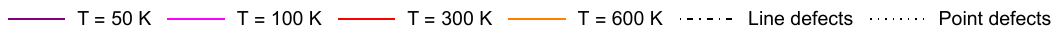}
    \caption{Imaginary part of the dimensionless longitudinal self-energy $\Sigma_L$ (left column) and transverse self-energy $\Sigma_T$ (right column) in SiO${}_2$ for the three values of the microscopic length cutoff $a$ considered in the main text. Top row: $a = 0.45$ nm. Middle row: $a = 0.5$ nm. Bottom row: $a = 0.55$ nm. }
    \label{fig:Ioffe-Regel-SiO2}
\end{figure*}

In the low frequency limit, we may thus approximate the imaginary part of both the longitudinal and transverse self-energies as
\begin{equation}
    {\rm Im} \Sigma = \frac{2 + \gamma^{-3} }{12 \pi} \frac{ n_P M^2}{ c_T^3 \rho^2 } \omega^3 \, .
\end{equation}
It is comforting to observe that, upon identifying $a^3 = M/\rho$, this is the same result obtained by Klemens\cite{Klemens1955} after averaging over phonon polarizations and introducing a defect density $n_P$.

Looking forward, there are two aspects of this simple model that deserve further discussion, and could be improved:
\begin{enumerate}
    \item At face value, Eqs.~\eqref{eq:sigmaTpointsApp} and~\eqref{eq:sigmaLpointsApp} imply a modification to the speed of sound in the material, as replacing $\omega = \omega(k)$ and taking $k \to 0$ yields a nonzero real number for the value of the self-energies. However, as we discussed earlier, there is no reason to have $r_0 = 0$ in general. And, in fact, for any value of $r_0 \neq 0$, no matter how small, the correction to the speed of sound due to point defects vanishes. Then, assuming that scattering by these point defects is only dominant at nonzero frequency --- in particular, when $r_2 \omega^2 \gg r_0 f^2$ --- we may safely neglect its effect. Testing these assumptions, and assessing whether a nonzero value of $r_0$ (and $r_4$, for that matter) can have a significant impact in our results, is left to future work.
    \item Strictly speaking, the Lagrangian in Eq.~\eqref{eq:point-defect-lagrangian} introduces a double-counting issue: in principle, it is not obvious how to distinguish between the continuum degrees of freedom in $u_i$ and the point defects in $Y_i$. Here we have introduced a somewhat arbitrary separation between the two, by postulating different dynamics for each of them. If we wanted to be fully consistent with the number of vibrational degrees of freedom one could obtain by counting the number of SiO${}_2$ units, we should make sure that the total phase space volume in our chosen degrees of freedom matches this value. These considerations may become relevant if the number density of defects $n_P$ becomes large enough that double-counting could become a quantitative issue. To make sure this is not an issue, one should come up with a Lagrangian describing point impurities that is directly given in terms of $u_i$, without introducing additional degrees of freedom to describe the effects of disorder. We also leave this to future work.
\end{enumerate}

\section{The Ioffe-Regel limit} \label{app:Ioffe-Regel}

To quantify whether the degrees of freedom in the elastic continuum description in our model are actually propagating modes, we take a closer look at the magnitude of the imaginary part of the self-energy $\Sigma_{T/L}$ for phonon propagation. Concretely, quasiparticles are no longer clearly defined if the imaginary part of the self-energy grows comparable to the real part, which sets the ``free'' dispersion relation. In the way we have defined it, the appropriate comparison to make is whether ${\rm Im} \{ \Sigma_{T/L} \}$ is smaller or larger than unity. We may do this on a case-by-case basis for all of the results we present in this work.

In Figure~\ref{fig:Ioffe-Regel-Glycerol}, we do so for the three fits to Glycerol data that we presented in the main text. The different values of $g_{\rm Glycerol}$ within each plot reveal that the microscopic physics picture in each case can be somewhat different: at smaller values of the coupling (with the length cutoff adjusted to match data) the imaginary part of the self energy is larger, being above $1/2$ for a broad frequency interval for $g = 0.994$ in the transverse mode. On the other hand, the imaginary part of the self-energy in each case is well-separated from unity for the largest value of the coupling. These features are all independent of temperature, and are a direct consequence of the string-phonon coupling, with no other interactions at play.

In Figure~\ref{fig:Ioffe-Regel-SiO2}, we do so for the three fits to SiO${}_2$ data that we presented in the main text, and within each plot we show the imaginary part of the self energy for different temperatures. What effectively varies among the different rows (different values of the microscopic length cutoff $a$) is the contribution from point defects, as the phonon-string coupling is uniquely fixed via the physical speed of sound. For the longitudinal self-energies, the imaginary part reaches a maximum of about $0.2$, well separated from unity. In the transverse case, this number is about $0.3$, which we also regard as well separated from unity. Interestingly (and reassuringly), the frequency at which phonons are impeded most strongly is controlled by the string-phonon coupling, and is virtually unaffected by point impurity scattering because the frequency is not high enough for it to become relevant. In this sense, point impurities act together with the UV regulator $a$, but leave the physics of long-wavelength phonon propagation unaffected.

In short, our formulation does not reach the Ioffe-Regel limit. That is, for the materials and temperature ranges we have considered, a description of thermal conductivity can be achieved \textit{without assuming} a ballistic approximation
in terms of
``propagons'' (our phonons) interacting with ``locons'' (our strings) without bringing ``diffusons'' (whose description in a continuum approach involving interaction of phonons  with elastic strings can be found in \cite{Churochkin2022}) into play.

\bibliography{main.bib,string_glass_biblio_3.0}

\end{document}